\documentclass[10pt]{article}
 
\usepackage{nonfloat}
\usepackage[affil-it]{authblk}
\usepackage{multicol}
\usepackage{graphicx}
\usepackage{amsmath}
\usepackage{amssymb}
\usepackage{cancel}
\usepackage{caption}
\usepackage{capt-of}
\usepackage{bbold}
\usepackage{xcolor,colortbl}
 
\usepackage[top=1.5in, bottom=1.0in, left=0.5in, right=0.5in]{geometry}

\newcommand{\argmax}{\arg\!\max}

\begin{document}

\vspace{-4cm} 

\title{Variance Dynamics - An empirical journey}
\author{Florent S\'egonne\thanks{\texttt{Electronic address: florent.segonne@80gmail.com}\\ \indent The author would like to thank the 80 Capital quantitative research team for numerous fruitful discussions, especially Tom Beyo, Lucas Plaetevoet, and Mindaugas Juozapavicius. Florent Bersani provided valuable comments and suggestions.
}
}
\affil{80 Capital LLP}
\date{}
\maketitle
\vspace{-1cm} 
 
\begin{center}
\line(1,0){540}
\end{center}
{\it \noindent We investigate the joint dynamics of spot and implied volatility from an empirical perspective. We focus on the equity market with the SPX Index our underlying of choice. Using only observable quantities, we extract the instantaneous variance curves implied by the market and study their daily variations jointly with spot returns. We analyze the characteristics of their individual and joint densities, quantify the non-linear relationship between spot and volatility, and discuss the modeling implications on the implied leverage and the volatility clustering effects. We show that non-linearities have little impact on the dynamics of at-the-money volatilities, but can have a significant effect on the pricing and hedging of volatility derivatives.}
\begin{center}
\line(1,0){540}
\end{center}

\begin{multicols}{2}
\section{Introduction}
Equity implied volatility, as priced by the market through vanilla options and volatility derivatives, is certainly not a constant. Its behaviour is strongly linked to its underlying, often appearing negatively-correlated with spot returns. Yet, at times, it exhibits spurs of independance, 
behaves capriciously, and displays a life of its own. Understanding the behaviour of volatility, be it for the purpose of risk management or the pricing and hedging of derivatives, is crucial to most market participants; unexpected moves can prove costly. \\\\
And the task is not easy. The meaning of implied volatility is rich as it inherently refers to multiple connected concepts; e.g. an at-the-money (ATM) implied volatility observed for a specific expiry on the entire volatility surface, or the full term-structure of variance. Its dynamics are complex and give birth to a range of distinctive regimes~\cite{gatheral-book}. 
It has led practitioners to define sets of rules to identify and trade around specific volatility patterns; among those, the concept of sticky-strike and sticky-delta~\cite{derman-99}, or the shadow gamma.  \\\\
%
%
%
%
Over the years, a host of volatility models have been introduced to better understand its complex behaviour. 
Those fall broadly into a few representative classes, of which pure stochastic volatility models (e.g. SABR~\cite{hagan-2002}, Heston~\cite{Heston93aclosed-form}, variance curve models~\cite{buehler-2006,bergomi-smiledynamicsII}) and spot-only-driven models (e.g. GARCH models~\cite{bollerslev-2008,engle-2008}) are probably the best known and most used. 
Recent advances have shown that the dynamics they generate are intrinsically constrained by the class they belong to~\cite{backus-1997,bergomi-smiledynamicsIV,ciliberti-bouchaud-potters-leverage-08,bergomi-guyon-2012,vargas-2013}. For instance, 
Bergomi demonstrates in~\cite{bergomi-smiledynamicsIV} that the dynamics of ATM implied volatilities generated by a stochastic volatility model are inherently linked to the smile produced by the model. Therefore, a volatility model often dictates more than the obvious, and should always be selected based on a clear understanding of the properties one wishes to model.\\\\
In this work, we study the joint dynamics of spot and implied volatility from an empirical perspective. Our journey into the volatility lanscape is pragmatic. We analyze the properties of volatility using as few assumptions as possible. Our aim is to identify and quantify the meaningful patterns of spot and implied volatility, and study their implications on the modeling of volatility. We proceed as follow.\\
We extract using only observable quantities the joint variations of the underlying market with the term-structure of implied variance (sect.~\ref{models}). Although direct observation of the term-structure is not possible, estimation with minimal distortion is achieved through the use of a general stochastic volatility framework (sect.~\ref{VolatilityModel}).\\ 
We then review step by step the model assumptions and discuss the limitations of our approach (sect.~\ref{adequacy}). The in-depth analysis serves as the basis to explore important properties of the spot/vol dynamics. 
We probe the characteristics of individual and joint densities and quantify the non-linear relationship between spot and volatility. We find that the implied leverage effect (i.e. the tendancy of atm volatility to increase as the underlying market decreases) and the volatility clustering effect (i.e. the propensity of volatility to stick to recent past levels, also referred to as the heterocedasticity of volatility) are well-captured by the combination of non-linear and non-Gaussian properties (sect.~\ref{sdf} and~\ref{jdf}). We then study the mean-reverting nature of volatility of volatility; although our goal is not to introduce another volatility model, we suggest some potential venues for improvements (sect.~\ref{VoV}). \\
In the last part, we gauge the impact of non-linearities on the pricing, modeling, and hedging of derivatives. We find them to have little influence on the dynamics of ATM volatilities:  on equity indices, the linear spot/vol correlation remains the dominant factor (sect.~\ref{skewnessskew} and~\ref{SSR}). However, convex effects change the realised volatility of annualised variance, thereby impacting the pricing and hedging of volatility derivatives (sect.~\ref{VVS}). Section~\ref{conclusion} concludes.
%
%
%
%
%
\section{Extracting short-term dynamics}
\label{models}
Focusing on short-term dynamics, we extract from observable quantities the daily variations of an asset $S_t$ and of the term-structure of its implied volatility $\xi_t^u$ for $u\geq t$. Our observables are listed futures on spot and on implied volatility.  Futures on implied volatility are a rich source of information on the term-structure of volatility, as long as one is careful enough to  correct for the small convexity adjustment present in those. In this paper, our example of choice is the SPX index and its corresponding volatility metric the VIX index\footnote{The term-structure of variance could also be extracted from the quotes of vanilla options using the well-known replication of variance. However, as attractive as it sounds, this approach is not trivial: it necessiates a complete option database and it also raises some non-trivial modeling questions, such as how to interpolate in time (between expiries) and in space (between strikes), or how to handle missing or incorrect data.}
%
%
\subsection{Dataset and notations}
The VIX index is a real-time measure of the market's implied variance of the SPX index over the next 30-calendar-days. For each trading time $t$, several futures are quoted. We denote them by $\mathcal{V}_t^{T_i}$ where $T_i$ is the expiration date - by an obvious extension, $\mathcal{V}_t^t$ represents the VIX index. These futures span a term-structure of several months, providing an observable but indirect measure of the implied variance priced by the market. \\\\
Although the trading in VIX futures began on March 26, 2004, liquidity remained low until 2008. The credit crisis changed the whole landscape. With volatility jumping suddenly to unexpected highs
, more and more market participants started to envision volatility as a potential hedge for their portfolio. In the years that followed, VIX trading increased significantly. Since 2012, approximately fifty thousand futures contracts trade on a daily basis on the first and second expiries (see Fig.~\ref{fig:vixvol}); medium-term futures, of approximatively 6-month maturity, can be traded with a reasonable liquidity, .e.g approximately five-thousand contracts per day.\\
\begin{minipage}{\columnwidth}
\includegraphics[width=\columnwidth]{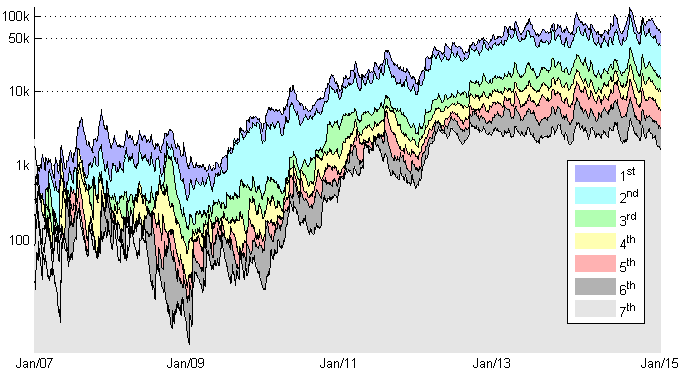}
\captionof{figure}{\it {\bf Liquidity of VIX Futures} Each curve represents in log-scale the average daily volume of the traded $n^{th}$ futures. The daily volumes have been averaged with an exponential kernel with half-life of one month.\\}
\label{fig:vixvol}
\end{minipage}
Our dataset comprises of the VIX index and of the VIX futures from the end of 2007. We exclude the early days of VIX trading; the lack of liquidity and the inconsistency of the quotes cannot be trusted. We focus on daily close-to-close variations. Consequently, our dataset contains more than 1500 daily observations, each observation point consisting of the VIX index and 7 futures.\\\\
%
%
%
%
We now introduce our notations. The risk-neutral market measure is denoted by $E_t^\mathfrak{M}[.]$, or simply by $E_t[.]$ when there is no ambiguity. The real-but-unknown measure is $E_t^\mathfrak{R}[.]$ that we approximate by the historical measure.
%
%
We denote by $\sigma_t^2 \delta t$ the variance realised by the spot process $S_t$ during times $t$ and $t+\delta t$, and $\text{var}^{T_1\to T_2}$ is the total variance realised during $T_1$ and $T_2$
\[
\text{var}^{T_1\to T_2}=\sum_{T_1}^{T_2}{\sigma_u^2 \delta u}\approx\int_{T_1}^{T_2}{\sigma_u^2 du}.
\]
We make a clear distinction between an achievable finite sampling, denoted by $\delta t$, and its theoritical limit, representing an infinitely-small instantaneous sampling, denoted by $dt$ and used profusely in stochastic calculus. The continuous formalism has the great advantage of simplifying proofs and equations, but can sometimes hide subtle sampling effects. The different integrals appearing in the text should often be interpreted as discrete sums over the discrete sampling period\footnote{That being said, we also approximate discrete sums by their equivalent integrals when this makes sense. For instance, sums such as $\sum_{i=1}^{N}{e^{-k (i-1)\delta t}\delta t}$ would be approximated by their integral counterpart $\int_t^T{e^{-u}du} = \frac{1-e^{-kN\delta t}}{k}$, valid as long as $\delta t$ is small or equivalently $N$ large. As we just said, the integral approximations have the advantage of greatly simplifying the equations.}. Because we are working with daily observations corresponding to trading bussiness days, the sampling frequency is by convention set to 252, i.e. $\delta t=\frac{1}{252}$. Note that the variance $\sigma_t^2$ is not known at time $t$, as it depends on the future return $\frac{\delta S_t}{S_t}=\frac{S_{t+\delta t}-S_t}{S_t}$ realised between $t$ and $t+\delta t$. \\\\
The time-$t$ term-structure of variance $\xi_t^u=E_t[\sigma_u^2]$ is not directly observable, but will be deducted from the values $\mathcal{V}_t^{T_i}$ of VIX futures. Because the instantaneous implied variance is a martingale under the risk-neutral measure~\cite{bergomi-smiledynamicsII}, we also have $\xi_t^u=E_t[\xi_u^u]$. This martingale property is the basis of numerous stochastic volatility models~\cite{bergomi-smiledynamicsII,buehler-2006,vargas-2013} ; We also follow that choice, because of the generality and flexibility of the resulting volatility models. \\\\
%
%
%
%
%
We denote by $\mathbb{V}_t^{T_1\to T_2}$ the fair-value at time $t$ of the annualised variance defined over the time interval $[T_1, T_2]$: 
\[
\mathbb{V}_t^{T_1\to T_2}=E_t[\frac{1}{T_2-T_1}\text{var}^{T_1\to T_2} ]=\frac{1}{\Delta T}\int_{T_1}^{T_2}{\xi_t^u du}
\]
When $t \leq T_1$, the forward-starting variance strike $\mathbb{K}_t^{T_1\to T_2}$ is equal to $\sqrt{\mathbb{V}_{t}^{T_1\to T_2}}$. \\\\
By definition, the level of the VIX index, also referred to as VIX cash or VIX spot, calculated at time $T_1$ should be equal to $\sqrt{\mathbb{V}_{T_1}^{T_1\to T_2}}$, where $T_2 = T_1+\frac{30}{365}$. In practice, the VIX is not exactly the 30-day implied volatility. First, it is necessary to take into account the correct number of returns in the next 30-day period by scaling the observed levels by a factor $\sqrt{\frac{252}{365}\frac{30}{\#\text{returns}}}$. In our dataset, we adjust each observation, i.e. the values of VIX index and VIX futures, by the correct factor without mentioning it explicitly. Second, it is a well-known fact that the VIX index, being computed as a linear interpolation between two incomplete strips of options, is only an imperfect proxy of variance. In particular, the VIX index exhibits two quirks, which can sometimes be misinterpreted for a real volatility impact. The first one is linked to the roll mechanism of the listed options, which can cause an artificial change in the VIX level (the historical methodology meant that 8 days before the expiry, the selected options roll from the $1^{st}$ and $2^{nd}$ expiries to the $2^{nd}$ and $3^{rd}$; with the emergence of weekly options, this issue has become less releavant). The second quirk comes from the fact that the addition (new trade) or deletion (trade closing) of some out-of-the-money options can lead to a sudden jump of the VIX value. Fortunately, VIX futures do not suffer from those artefacts, as they are only an expectation of the future VIX index.\\\\
With the previous notations, we can finally express the value at time $t$ of a VIX future expiring at maturity $T_1$:
\begin{eqnarray}
\label{eq:vix-definition}
\mathcal{V}_t^{T_1} = E_t[\sqrt{\mathbb{V}_{T_1}^{T_1\to T_2}}]=E_t[\sqrt{\frac{1}{\Delta T}\int_{T_1}^{T_2}{\xi_{T_1}^u du}}]
\end{eqnarray}
%
%
 
\subsection{Spot model}
The modeling of the spot returns is rather natural. We model the dynamics of the return $r_t=\frac{\delta S_t}{S_t}$ as a stochastic realisation of the instantaneous implied variance $\xi_t^t$ over the finite time interval $\delta t$. Working with futures written on the SPX index, the spot model is defined as follow:
\begin{eqnarray}
\frac{d S_t}{S_t} = \sqrt{\xi_t^t} d Z_t,
\label{eq:spot}
\end{eqnarray}
where $d Z_t$ is a centered random variable of variance $dt$.  Consequently, the equality $\xi_t^t = E_t[\sigma_t^2] = E_t[\frac{1}{\delta t}(\frac{\delta S_t}{S_t})^2]$ is naturally enforced. \\\\
Although the variable $\frac{1}{\sqrt{\delta t}}\delta Z_t$ is centered and normalized to unity under the risk-neutral measure, it is not necessarily in practice. We denote by $\mu_Z$ and $\sigma_Z$ the annualised trend and volatility of the stochastic process $Z_t$ under $\mathfrak{R}$: 
\[
\mu_Z \delta t = E_t^\mathfrak{R}[\delta Z_t] \ \text{and} \ \sigma_Z^2 \delta t = E_t^\mathfrak{R}[(\delta Z_t-\mu_z\delta t)^2]
\] 
This discrepancy between market view and realised variance is the basis of the volatility risk premium : the predictive power of the implied instantaneous variance is quite poor, and for most times, $\xi_t^t \geq E_t^\mathfrak{R}[\sigma_t^2]$, or equivalently $\sigma_Z \leq 1$ (see section~\ref{sdf}).\\\\
Finally, note that the variable $\delta Z_t$ does not need to be Gaussian, allowing the modeling of the discrete nature of the returns. We denote by $\zeta = E^\mathfrak{R}[\delta Z^3]$ and $\kappa=E^\mathfrak{R}[\delta Z^4]-3$ its skew and excess kurtosis. 
%
%
%
\subsection{Stochastic variance model}
\label{VolatilityModel}
Our goal is to extract the factors driving the term-structure of variance $\xi_t^u$ as accurately as possible and without having to rely significantly
on any volatility model. However, as can be seen from Eq.~\ref{eq:vix-definition}, a VIX future provides only an indirect measure of the implied variance curve between times $T_1$ and $T_2$. The lens of observation is the risk-neutral market pricing, which must be modeled in order for us to reach back the underlying variable $\xi_t^u$. Therefore, we must select a volatility model capable of pricing VIX futures. \\\\
Selecting a volatility model allows us to derive, for each observation date $t$, an accurate estimate of the term-structure of volatility $\xi_t^u$ implied by observable VIX futures $\mathcal{V}_t^{T_1}$. This is achieved through a pricing equation (Eq.~\ref{eq:vix}) that defines the convexity correction inherent to VIX futures. The convexity adjustment is a function of the model parameters, which are estimated from the daily variations of VIX futures (through Eq.~\ref{eq:variation}). Although distinct volatility models would lead to different convexity adjustments, the differences would be small and would not change the analysis. In practice, the magnitude of the convexity correction is so small that the convexity adjustment is almost inconsequential for our purpose. 
%
Once calibration is achieved, we will be in a position to revisit and discuss some of our modeling assumptions (sections~\ref{analysis} and~\ref{consequences}).\\
\begin{minipage}{\columnwidth}
\vspace{0.25cm}
\includegraphics[width=\columnwidth]{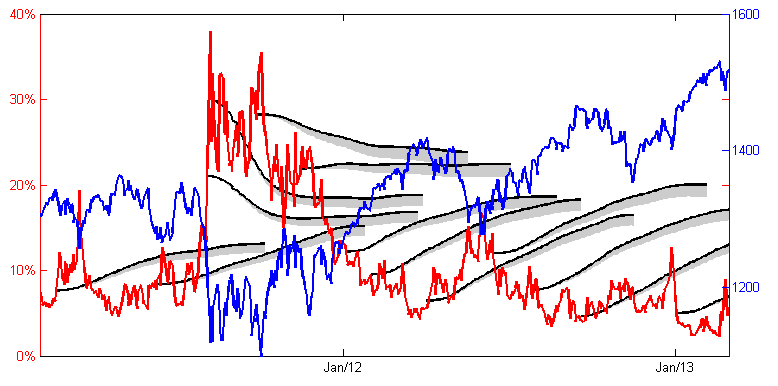}
\captionof{figure}{\it {\bf Instantaneous Variance Curves} We graph some estimated instantaneous variance curves (in black) over the period $2011-2012$, as well as the corresponding magnitude of the convexity correction in shaded gray below each curve. The VIX cash index is displayed in red and the underlying market in blue.}
\label{fig:curves}
\end{minipage} 
\subsubsection{Description of the volatility model}
The rather-unpredictable nature of volatility is usually modeled using stochastic models\footnote{The use of GARCH models would not be appropriate in our case, as they do not possess any volatility factors. They ignore the independant nature of volatility, which is exactly what we aim to model. As such, they would be too restrictive for our purpose. It is interesting to note that by specifying the stochastic volatility factors as deterministic functionals of the spot factor, some GARCH models can be interpreted as reduced-versions of more-general stochastic volatility models - see section~\ref{garch}}. Quite generally, the instantaneous term-structure of variance $\xi_t^u \text{ with } t\leq u$ is assumed to be driven by a set of $n$ Brownian motions:\\
\vspace{-0.5cm}
\begin{eqnarray}
\label{eq:var-model}
d \xi_t^u = \hspace{0.5cm} \xi_t^u\times \sum_{\alpha=1}^{n} {\theta_\alpha \omega^{\alpha}(t,u,\xi_t) d W^{\alpha}_t}
\end{eqnarray}
with correlation matrix $C_{\alpha,\beta}= \frac{1}{dt}\left < dW_t^\alpha , dW_t^\beta \right> =  \rho_{\alpha,\beta}$. \\\\
Although the stochastic variance model is expressed in a continuous setting, it is really a discrete framework that is described by the above equation - most often than not, the infinitesimal term $d$ should be understood as a finite variation $\delta$. The above framework, which is built on the martingale properties of implied variance, is quite general and flexible. Our choice of using a log-normal model was motivated by recent studies~\cite{gatheral-2008}, as well as its popularity. Working at a finite time scale, non-linearities can easily be taken into account by simply introducing non-linear relationships between volatility factors and spot returns (see section~\ref{jdf}). \\\\
We denote by $\Theta$ the diagonal matrix with diagonal terms $\theta_\alpha$ and $\Omega$ the covariance matrix defined by $\Omega_{\alpha,\beta}=\theta_\alpha \theta_\beta\rho_{\alpha,\beta}$ - the coefficients $\theta_\alpha$ represent the volatility of the $\alpha$ factors. The instantaneous volatility $\nu_{t}^{u}$ of the instantaneous variance is maturity-dependant and equal to:
\begin{eqnarray}
\label{eq:varVoV}
\nu_{t}^{u}=\sqrt{\sum_{\alpha,\beta} {\Omega_{\alpha,\beta}\omega^{\alpha}(t,u,\xi_t)\omega^{\beta}(t,u,\xi_t) }}
\end{eqnarray}
The weighting functions $\omega^{\alpha}$ might depend on the curve $\xi_t$ and time, but not on the spot $S_t$ - this is the choice followed in~\cite{bergomi-smiledynamicsIV,buehler-2006,vargas-2013}. Quite often, they are chosen as time-invariant decreasing functions, i.e. $\omega^{\alpha}(t,u,\xi_t)=\omega^{\alpha}(u-t,\xi_t)$, expressing the fact that a random shock at time $t$ impacts the whole term-structure of variance with a magnitude $\omega^\alpha(u-t,\xi_t)$ decreasing with maturity $u-t$. Frequently, the functions are defined as exponentials $\omega^\alpha(t,u,\xi_t)=\exp{(-k_\alpha(u-t))}$. As a result, the stochastic model becomes Markovian and can be integrated exactly in closed-form~\cite{bergomi-smiledynamicsII}. Although we do not need these explicit properties, the additive separability of exponentials greatly simplifies the analysis. We follow that choice in our numerical simulations.\\\\
Equations~\ref{eq:spot} and ~\ref{eq:var-model} defines a general stochastic spot-vol model. When $\theta_\alpha = 0$, the model reduces to a simple Black-Scholes (BS) model with a deterministic, spot-independant, diffusion variance defined by $\forall u\geq t, \ \xi_u^u=\xi_t^u$. In this case, the Black-Scholes volatility is also the variance-swap volatility $\sigma_{VS}(t,T) = \sqrt{\mathbb{V}_{t}^{t\to T}}$, the volatility smile being obviously flat.\\
In practice, a small number of driving factors is sufficient to accurately capture the dynamics of variance. Cont and da Fonseca show that the 3 principal modes represents $98\%$ of the variance of the daily curve deformations~\cite{cont-fonseca-01}. The first mode, amounting to $80\%$ , can be interpreted as a level effect, whereas the second and third modes correspond respectively to slope and convexity. Our data set does not capture the long-end of the variance curve since every observation is limited to the first seven expiries, just above half a year. As such, using a large number of factors might lead to overfitting and parameter instabilities. For that reason, we follow the methodology introduced by Bergomi in~\cite{bergomi-smiledynamicsII} and Gatheral in~\cite{gatheral-2008}, and select two factors only (note that we also investigated the use of a three-factor model, but did not observe any significant improvement - see discussion insection~\ref{adequacy}).\\\\
%
%
%
%
%
%
%
%
%
From Eq.~\ref{eq:vix-definition} and~\ref{eq:var-model}, we can derive the time-$t$ value of a VIX future $\mathcal{V}_t^{T_1}$, as well as its variation $d\mathcal{V}_t^{T_1}$. In particular, a VIX future is always below\footnote{\label{ftnt:vixarb}Although VIX futures should always quote below their corresponding forward-variance level, it is not always the case in practice. Dislocations do appear from time to time. However, those are extremely hard to capture, as bid-offers render the arbitrage impossible. Those dislocations are not frequent, and would not change the results of our analysis.} the corresponding forward-starting variance 
\[
\mathbb{K}_t^{T_1}=\mathbb{K}_t^{T_1\to T_2}=\sqrt{\mathbb{V}_{t}^{T_1\to T_2}}=\sqrt{\frac{1}{\Delta T}\int_{T_1}^{T_2}{\xi_{t}^u du}}
\]
The convexity adjustment comes from the concavity of the square-root function and is proportional to the vol of vol parameters $\Omega_{\alpha,\beta}$. \\\\
Introducing the notation $\mathbb{K}_t^{T_1,\alpha} = \sqrt{\frac{1}{\Delta T}\int_{T_1}^{T_2}{\xi_t^u \omega^{\alpha}(t,u,\xi_t)}du}$, we finally obtain the set of equations that we will use throughout the paper (see appendix~\ref{app:convexitycorrection} for the derivation steps):\\
\vspace{0.4cm}
{
\hspace{-0.5cm}
\fbox{
\hspace{-0.5cm}
\begin{minipage}{\linewidth}
\vspace{-0.035cm}
\begin{eqnarray}
&&\text{\underline{Pricing Equation}} \label{eq:vix} \\
\mathcal{V}_t^{T_1} &=&  \mathbb{K}_t^{T_1}\times (1-\text{convexity correction})\nonumber \\
\text{c.c.}            &=& \sum_{\alpha,\beta} {\frac{\Omega_{\alpha,\beta}}{8} \frac{e^{(k_\alpha + k_\beta)(T_1-t)}-1}{k_\alpha + k_\beta}  \frac{(\mathbb{K}_t^{T_1,\alpha} )^2(\mathbb{K}_t^{T_1,\beta})^2}{(\mathbb{K}_t^{T_1})^{4}}}\nonumber  \\
\label{eq:variation}
&&\text{\underline{Model Dynamics}}\\
\frac{d \mathcal{V}_t^{T_1}}{\mathcal{V}_t^{T_1}} &=&  \sum_\alpha {\frac{\theta_\alpha}{2} (\frac{\mathbb{K}_t^{T_1,\alpha}}{\mathcal{V}_t^{T_1}})^2  d W^\alpha_t}\nonumber \end{eqnarray}
\end{minipage}
}}
In the remainder of this section, we describe the calibration of our volatility model and the extraction of the volatility factors. The reader uninterested by the technical details can safely jump to section~\ref{analysis}. 
%
%
\subsubsection{Fitting the volatility model}
The stochastic model is specified by the set of parameters $\Xi$, which comprises of  the number of Wiener processes $W^\alpha$ driving the instantaneous variance (set to two in this work), of the shape of the kernel functions $\omega^\alpha$ (defined by the parameters $k_\alpha$), and of the corresponding covariance structure  $\Omega_{\alpha,\beta}$.\\\\
Our volatility model, which aims at capturing the short-term variations of VIX futures $\frac{\delta \mathcal{V}_t^{T_i}}{\mathcal{V}_t^{T_i}}$, is not perfect and will fail to match perfectly the variations of the entire term-structure. This is partly due to the inadequacy and simplicity of our model, but not only. Due the complexity of volatility dynamics, any volatility model would fail at some levels and some matching errors will always be present. Those would be also enhanced (and sometimes caused) by the illiquidity of some VIX futures and the inacuraries of their quotes.\\\\
These matching errors must be accounted for. To do so, we follow a standard approach and introduce for each variation $\frac{\delta \mathcal{V}_t^{T_i}}{\mathcal{V}_t^{T_i}}$ a measurement error-term $\eta_t^{i}$. The measurement term is modeled as Gaussian noise with volatility $\sigma_t^{L,i}$ inversely proportional to the current liquidity of the corresponding VIX futures. By doing so, we force the variations of the most liquid futures, i.e. the ones that
are traded the most, to be better modelled by our volatility model than less-liquid (e.g. longer-
term) futures\footnote{Note that our dataset also includes VIX index levels that do not possess any liquidity since the VIX index is not tradeable. In addition, we have seen that the VIX index is more prone to unacceptable variations (by construction). To alleviate these issues, we do not use the the VIX index variations to calibrate our model parameter (in Eq.~\ref{eq:variation}). However, we do use its value to extract more accurately the term-structure of variance $\xi_t$ from Eq.~\ref{eq:vix}.}. Consequently, a future's variation is the sum of a model-term and an error-term:
\begin{eqnarray}
\label{eq:matrix}
\frac{\delta \mathcal{V}_t^{T_i}}{\mathcal{V}_t^{T_i}} = \underbrace{ \sum_\alpha {\frac{\theta_\alpha}{2} (\frac{\mathbb{K}_t^{T_1,\alpha}}{\mathcal{V}_t^{T_1}})^2  \delta  W^\alpha_t}}_{\text{model-term}} + \underbrace{\sigma_t^{L,i} \times \eta_t^{i}}_{\text{error-term}}.
\end{eqnarray}
We introduce the additional notations:\\
\hspace{1cm} - ${D_t}$ is a diagonal matrix ${{D_t}}(i,i)=\sigma_t^{L,i}$\\
\hspace{1cm} - $M_t$ is the matrix defined by ${M_t}(i,\alpha) = \frac{1}{2} (\frac{\mathbb{K}_t^{T_i,\alpha}}{\mathcal{V}_t^{T_i}})^2$\\
\hspace{1cm} - $\delta W_t$ is the Gaussian column-vector with components $\delta W_t^\alpha$\\
\hspace{1cm} - $U_t$ is the normalized Gaussian $U_t = \frac{1}{\sqrt{dt}}\text{TrI}^{-1}\delta W_t $ with $\text{TrI}$ a lower triangular matrix such that $\text{TrI}\times\text{TrI}^\top=C$ (i.e. Cholesky's decomposition). Working with $U_t$ or $\delta W_t$ is equivalent, but $U_t$ has the property of having uncorrelated normalized-to-unity components.\\\\
This allows us to recast the above equation in matrix form as:
\begin{eqnarray}
\frac{\delta \mathcal{V}_t}{\mathcal{V}_t} = M_t \times \underbrace{\Theta \times \text{TrI} \times U_t \times \sqrt{dt}}_{\Theta \times \delta W_t} + {D_t} \times \eta_t
\end{eqnarray}
%
%
%
%
At this point we are ready to rephrase our calibration procedure into a Bayesian framework. One can express the probability of our observations as $p(\dots,\mathcal{V}_t,\dots|\Xi)$  and extract the model parameters by maximum-likelihood: 
\begin{eqnarray*}
\Xi^\star &=& \argmax_\Xi p(\dots,\mathcal{V}_t,\dots|\Xi)
\end{eqnarray*}
Although maximum-likelihood have notorious convergence problems (often due to the presence of numerous local minima), we did not experience such issue with two factors - however, with three factors, numerous optimizations from randomly selected random points had to be run.
The joint probability can be decomposed as a product of independant probabilities:
\begin{eqnarray}
p(\dots,\mathcal{V}_t,\dots|\Xi) &=& \prod_t { p(\frac{\delta \mathcal{V}_t}{\mathcal{V}_t}|\xi_t,\Xi)} \nonumber \\
                                &=& \prod_t { {\int_{\delta W_t} p(\frac{\delta \mathcal{V}_t}{\mathcal{V}_t},\delta W_t|\xi_t,\Xi)}}
                                \label{eq:step1}
\end{eqnarray}
Traditionally, the last integral cannot be integrated directly and one usually resorts to an iterative expectation-maximization algorithm thanks to a conventional Jensen argument. Fortunately for us, integrating Eq.~\ref{eq:step1} does not present any difficulty since the joint density $p(\frac{\delta \mathcal{V}_t}{\mathcal{V}_t},\delta W_t|\xi_t,\Xi)$ can be expressed as a simple product of Gaussian multivariates
\[
\underbrace{p(\frac{\delta \mathcal{V}_t}{\mathcal{V}_t}|\delta W_t,\xi_t,\Xi)}_{error}\times \underbrace{p(\delta W_t|\Xi)}_{prior}
\]
Using the well-known identity for the integration of a multivariate variable $x$ of dimension $n$
\[
\int_{-\infty}^{\infty}{\exp(-\frac{1}{2}x^\top A x + J^T x)dx}=\frac{(2\pi)^\frac{n}{2}}{|A|^\frac{1}{2}}\exp[\frac{1}{2}J^\top A^{-1}J]
\]
we find that the log of the integral $\int_{\delta W_t} p(\frac{\delta \mathcal{V}_t}{\mathcal{V}_t},\delta W_t|\xi_t,\Xi)$ is proportional (ignoring useless constant term) to:
\begin{eqnarray*}
&-&\log{|D_t|}-\log{|\text{Id}+\sqrt{\Omega}^\top M_t^\top D_t^{-1} M_t \sqrt{\Omega} dt|}\\
&+& (\mu + \sqrt{\Omega}^\top M_t^\top D_t^{-1} \frac{\delta \mathcal{V}_t}{\mathcal{V}_t} \sqrt{dt} )^\top \dots\\
&&\times\  (\text{Id}+\sqrt{\Omega}^\top M_t^\top D_t^{-1} M_t \sqrt{\Omega} dt)^{-1}\dots\\
&&\times \ (\mu + \sqrt{\Omega}^\top M_t^\top D_t^{-1} \frac{\delta \mathcal{V}_t}{\mathcal{V}_t} \sqrt{dt}) \\
&-& ( \frac{\delta \mathcal{V}_t}{\mathcal{V}_t})^\top D_t^{-1}  \frac{\delta \mathcal{V}_t}{\mathcal{V}_t} - \mu^\top \mu
\end{eqnarray*}
where $\mu = E_t^\mathfrak{R}[U_t]=E^\mathfrak{R}[U]$. Although the risk-neutral expectation of the factors $W_t^\alpha$ is zero (since the instantaneous variance is a martingale under the risk neutral measure), nothing guarantees that this property should also hold under the real-but-unknown measure. It is actually a well-known fact that the implied variance has realized negative decay, i.e. $E_t^\mathfrak{R}[\delta W_t]\leq0$, i.e. giving rise to the well-known term-structure volatility risk premia. \\\\
Solving for the model parameters is now straightforward. Because the matrices $M_t$ depend on the estimated curves $\xi_t$, which themselves depend on the set of parameters $\Xi$, we must proceed iteratively in pseudo expectation-maximization fashion:
%
%
%
%
\begin{enumerate}
\item {\it Modeling step} First, given a fully specified model (i.e. a full set of parameters $\Xi$), the instantaneous variance term-structure $\xi_t^u$ can be extracted from Eq.~\ref{eq:vix} for each time $t$. Without any assumption on the curve $\xi_t$, our problem would not be tractable. We assume that the variance term-structure is smooth\footnote{
A more realistic modelling of the term-structure of variance would only assume piece-wise smooth functions, with discontinuities happening at important financial dates (e.g. Federal Reserve board meetings, release of key indicators). This might prove crucial for stocks, but less for global indices. Note that instead of decomposing the curve $\xi_t$ onto a small set of basis functions, we could have instead added to the pricing equation~\ref{eq:vix} a Tychonoff regularization term. We experimented with both approaches and did not  notice any significant differences. 
} and parameterize each variance curve by a small number of basis functions capturing most of the variability of the curve. Figure~\ref{fig:curves} graphs some examples of estimated curves. 
\item {\it Expectation step} Once the curves have been estimated, the integrals $\mathbb{K}_t^{T_i,\alpha}$ appearing in Eq.~\ref{eq:variation} (or equivalently the matrix $M_t$ in Eq.~\ref{eq:matrix}) can be computed.
\item {\it Maximisation step}  We are then in a position to determine the unknown parameters by simple maximum-likelihood. In practice, to avoid convergence problems, the kernel parameters $k_\alpha$ are iteratively selected on a spanning grid, being kept fixed while the remaining parameters are estimated by maximum-likelihood.
\end{enumerate}
%
%
%
We iterate the above steps untill convergence has been achieved. On our dataset, the optimal set of parameters is provided in the below table.
\begin{center}
  \begin{tabular}{| c | c | c | c | c | c | c |}
    \hline
   $k_F$ & $k_S$ & $\mu_F$ & $\mu_S$ & $\theta_F$ & $\theta_S$ & $\rho$ \\ \hline
    $10.25$ & $1.05$ & $-7.5\%$ & $-0.4\%$ & $180\%$ & $92\%$ & $51\%$ \\
    \hline
  \end{tabular}
\end{center}
Bergomi's two factor model has similar parameters. However, in our calibration, the slow factor $k_S$ is significantly higher (approximately 1 to be compared with 0.35 in~\cite{bergomi-smiledynamicsII}), mainly due to the fact that our model is focused on short-to-medium-term maturities (there is no need to model the long-tend of the variance curve).\\\\
Once the optimization is finished, we extract the hidden states $\delta W_t$  by maximizing the posterior:
\[
p(\delta W_t | \frac{\delta \mathcal{V}_t}{\mathcal{V}_t},\xi_t,\Xi) \propto p(\frac{\delta \mathcal{V}_t}{\mathcal{V}_t}|\delta W_t,\xi_t,\Xi)\times p(\delta W_t|\Xi)
\]
The states $\delta W_t$ are the driving factors of implied volatility in our general stochastic volatility framework. They are a simplified representation of the implied volatility term-structure variations.\\
%
%
\begin{minipage}{\columnwidth}
\vspace{0.25cm}
\includegraphics[width=\columnwidth]{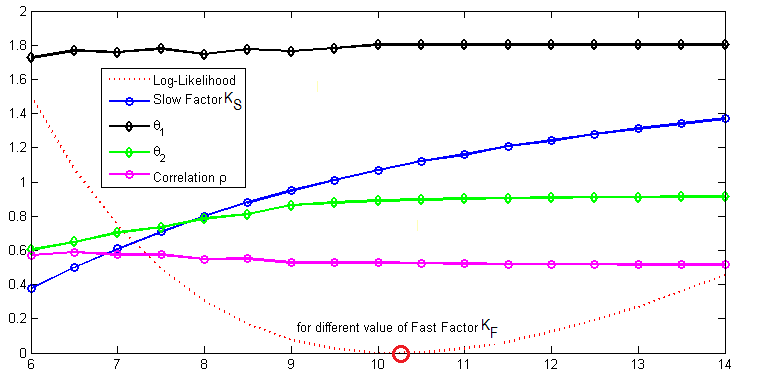}
\captionof{figure}{\it {\bf Optimization} We graph the optimized estimated parameters as a function of the selected fast factor $k_F$. The minimum of the log-likelihood (in dotted red) is achieved for a fast factor $k_F \approx 10.25$. For different values of $k_F$, the other parameters do not vary significantly. For instance, the estimated correlation $\rho_{\alpha,\beta}$ and the first variance parameter $\theta_F$, are quite stable around 0.5 and 1.8 respectively. The slow factor $k_S$, and the second variance parameter $\theta_S$ are slowly increasing as the parameter $k_F$ is increased.}
\label{fig:opt}
\vspace{0.25cm}
\end{minipage}
\subsubsection{Convergence and Stability}
Convergence is achieved in a couple of iterations. At each iteration, the expectation step might incur some errors (e.g. due for instance to incorrect model parameters). Those are unlikely to cause any significant inaccuracies in the estimation of the integrals $\mathbb{K}_t^{T_1,\alpha}$ - mainly because the convexity correction magnitude is quite small and could almost be ignored - see section~\ref{magnitude}. Figure~\ref{fig:curves} displays some of the estimated instantaneous variance curves along with the corresponding convexity corrections.\\\\
The accurate identification of the kernel functions is more difficult. Whereas it is clear that two distinct modes exist, i.e. one fast $k_F > 5$ and one slow $k_S<1.5$, the log-likelihood of different and apparently-acceptable solutions do not always differ significantly. For a wide range of acceptable solutions, encompassing the range $6<k_F<14$ and $0.3<k_S<1.4$, the remaining parameters $\theta_F,\theta_S,\rho$ are quite stable. Figure~\ref{fig:opt} illustrates this point.
\subsubsection{Orders of magnitude}
\label{magnitude}
We provide some ballpark numbers on the magnitude of the convexity correction. To do so, we make the common assumption of a relatively-flat term-structure of variance. For instance, focusing on the interval $[T_1,T_2]$, this assumption means that the difference between the instantaneous variance $\xi_t^u$ for $u\in [T_1,T_2]$ and the annualized variance $\mathbb{V}_{t}^{T_1\to T_2}$ is negligible compared to the variance itself $\mathbb{V}_{t}^{T_1\to T_2}$. Mathematically, this is often formulated as $\forall u\in [T_1,T_2], \ (\xi_t^u - \mathbb{V}_{t}^{T_1\to T_2})=\text{o}(\mathbb{V}_{t}^{T_1\to T_2})$. We will use this assumption several times throughout this work. \\\\
The integrals $\mathbb{K}_t^{T_1,\alpha}$ can be approximated at first-order by $\mathbb{K}_t^{T_1}e^{-\frac{1}{2}k_\alpha(T_1-t)}\sqrt{g(k_\alpha \Delta T)}$ with $g(x)=\frac{1}{x}\int_0^x{e^{-u}du}=\frac{1-e^{-x}}{x}$ and $\Delta T = T_2-T_1 = \frac{30}{365}$. 
From there, the convexity adjustment of Eq.~\ref{eq:vix} can also be approximated:
\[
\text{c.c.}  \approx \frac{1}{8}\left(\sum_{\alpha,\beta}{\Omega_{\alpha,\beta}\underbrace{g(k_\alpha \Delta T)}_{g_\alpha(\Delta T)}\underbrace{g(k_\beta \Delta T)}_{g_\beta(\Delta T)}\frac{1-e^{-(k_\alpha + k_\beta)(T_1-t)}}{k_\alpha + k_\beta}}  \right),
\]
which converges to a limit of $\frac{1}{8}\left(\sum_{\alpha,\beta}{\frac{\Omega_{\alpha,\beta}g_\alpha(\Delta T) g_\beta(\Delta T)}{k_\alpha + k_\beta}} \right)$ for long-term expiries. However, note that the limit for large maturities should not be trusted. Our model has been calibrated on short-term to medium-term expiries, and the simplicity of our model would not be appropriate to evaluate the longer-end of the curve (over 6-months). \\\\
Plugging our model parameters, we find that the convexity adjustment is quite small for short maturities. For the first two futures, the convexity is less than $5\%$, and could almost be neglected; the limiting convexity adjustment for long-term maturities is smaller than $10\%$.\\
\begin{minipage}{\columnwidth}
\includegraphics[width=\columnwidth]{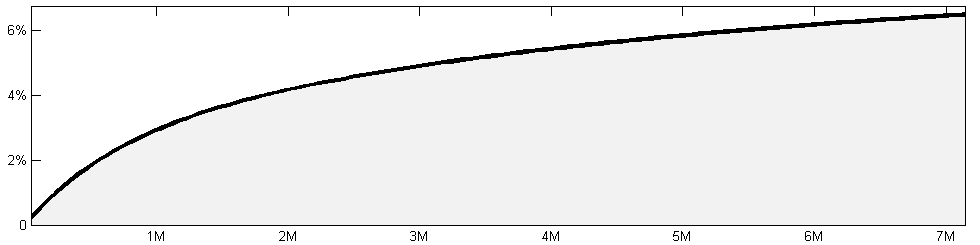}
\end{minipage}
Because the convexity adjustment can be neglected at first-order, we can assimilate the value of a VIX future to its corresponding forward-variance strike, i.e. $\mathcal{V}_t^{T_1}\approx \mathbb{K}_t^{T_1}$. We immediately deduct that the volatility of a VIX future can be approximated by:
\begin{eqnarray}
\label{eq:vixvov}
\frac{1}{2}\sqrt{\sum_{\alpha,\beta} {\Omega_{\alpha,\beta}g_\alpha(\Delta T)g_\beta(\Delta T)e^{-(k_\alpha+k_\beta)(T_1-t)}}},
\end{eqnarray}
which is slightly lower than half the volatility of variance $\nu_{t}^{T_1}$ as defined in eq.~\ref{eq:varVoV}.
%
%
%
%
%
%
%
%
%
%
%
\section{Analysis and Discussion}
\label{analysis}
At this stage, our volatility model has been calibrated, the model factors $\delta Z_t$ and $\delta W_t^\alpha$, as well as the daily variance curves $\xi_t$ have been estimated. 
Before proceeding any further, we check that our model can be trusted for the purpose of better understanding volatility. We verify that it accurately captures the main modes of variance curve deformations (section~\ref{adequacy}). This check validates the use of hidden states $\delta W_t$ to analyse the properties of volatility in a simpler low-dimensional setting. The volatility factors constitute a simplified representation of the term-structure variations.\\\\
We are then in a position to explore the joint dynamics of spot and volatility. We proceed step by step. 
We probe the characteristics of the individual densities of the model factors, discuss the validity of the Gaussian assumption and the implications on the magnitude of the volatility risk premia (section~\ref{sdf}). We then focus on the joint variations and model the non-linear relationship between spot and volatility (section~\ref{jdf}). We highlight the link with GARCH models, and study the impact on the implied leverage effect and the heterocedasticity of volatility. Finally, we delve into the volatility of volatility itself (section~\ref{VoV}). The mean-reverting nature of the VVIX index suggests some possible improvements.\\\\
\subsection{Model Adequacy}
\label{adequacy}
To evaluate the adequacy of the stochastic variance model with historical data, we conduct two elementary experiments. \\\\
To begin, we compare the theoritical term-structure of the VIX volatility (given by our calibrated model in Eq.~\ref{eq:vixvov}) with historical realized levels. As figure~\ref{fig:vixfit} illustrates the match is satisfactory. The limit for short-maturities, as $u-t\to 0$, can be computed around $90\%$, a value slightly below the observed volatility of the VIX index (at approximately $110\%$ since 2007). Figure~\ref{fig:vixfit} also shows that short-term volatility is more volatile and display more skewness and kurtosis than the long-end of the curve (see section~\ref{sdf}).\\
\begin{minipage}{\columnwidth}
\vspace{0.5cm}
\includegraphics[width=\columnwidth]{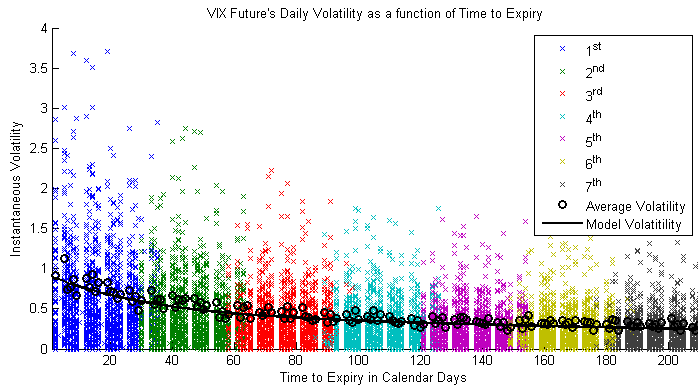}
\captionof{figure}{\it {\bf Instantaneous VIX volatility} Each cross corresponds to the realized daily volatility of a VIX futures plotted as a function of its time to expiry. For each given maturity, the circle represents the quadratic average of all corresponding daily volatilities. The curve represents the model volatility computed as in Eq.~\ref{eq:vixvov}.}
\label{fig:vixfit}
\vspace{0.5cm}
\end{minipage}
In a second experiment, we compute from the set of calibrated curves $\xi_t$ the principal orthogonal modes of the curve variations $\frac{\delta \xi_t}{\xi_t}$ (i.e. Karhunen-Lo\`eve decomposition). The first three principal eigenmodes captures more than $99\%$ of the total variance. Figure~\ref{fig:modes} displays the corresponding modes. The first mode, covering almost $95\%$ of the total variance, corresponds to a level effect, whereby the whole curve deforms subject to an implied volatility shock. The deformation is not uniform, but affects more the short-term portion of the curve, reflecting the higher volatility of short-term futures. As described by Cont and da Fonseca in~\cite{cont-fonseca-01}, the second and third eigenmodes can be identified to slope and convexity effects. \\\\
On this data set, the convexity effect provided by the third eigenmode is negligeable as its contribution to the total variance is less than $1\%$. This is explained by the short time span provided by the first 7 VIX futures. Two principal modes are sufficient to capture $99\%$ of the deformation modes. This explains why using more than two factors in our model calibration might lead to a difficult optimization and overfitting. To illustrate the adequacy of our simple model with the data set, we compute the two orthogonal modes implied by our stochastic model. Figure~\ref{fig:modes} displays the modes implied by the data (i.e. from estimated variance cuves) and the model. The match between the model and the data is surprisingly good, except maybe for very short-term maturities (less up to 2 weeks), where calibrated variance curves appear ``flatter'' than the exponential model modes. This might be due to the smoothing constraint that we introduced for the term-structure of variance. The emergence of short-term VIX futures, introduced in 2014 by CBOE, will progressively alleviate this issue.\\
\begin{minipage}{\columnwidth}
\vspace{0.5cm}
\includegraphics[width=\columnwidth]{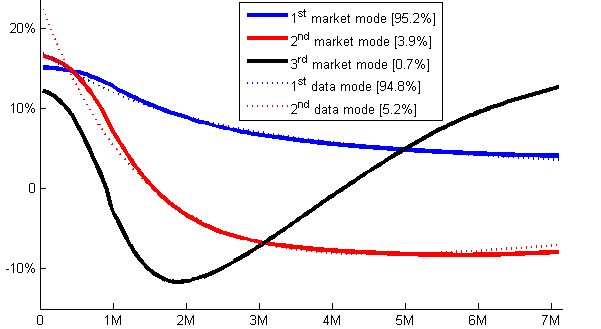}
\captionof{figure}{\it {\bf Market and Model Modes } The modes implied by the data and the model are display in solid and dotted line respectively. In both cases, the first mode represents $95\%$ of the total variance. The second mode corresponds to a slope term, whereas the third mode captures convexity. Two modes are sufficient to accurately captures $99\%$ of the variance.}
\label{fig:modes}
\end{minipage}
%
%
%
\subsection{The Gaussian is not the ``normal''}
\label{sdf}
We now turn our attention to the statistical properties of the stochastic factors $\delta Z_t$ and $\delta W^\alpha_t$, and of the variations of the instantaneous variance $\frac{\delta \xi^{t}_t}{\xi^{t}_t}$. We compute some elementary statistics over the whole period of study.  
\begin{center}
  \begin{tabular}{| c | c | c | c | c | c | c |}
    \hline                                                                                                
		X          & $\mu_X$                         & $\sigma_X$ & $\zeta_X$              & $\kappa_X$   & $\nu_X^+$ & $\nu_X^-$\\ \hline
    $\delta Z_t$                                   & $+33\%$                          & \color{red}{$79.6\%$} & $-0.57$           &             $1.59$                   & $5.23$ & $3.76$                                 \\ \hline
$\delta W^F_t$                                & \color{blue}{$-117\%$}             & $100\%$          & $+0.36$           &                $4.25$                                   & $3.25$ & $3.35$                                                            \\ \hline
 $\delta W^S_t$                                &  \color{blue}{$-68\%$}              & $100\%$          & $+0.43$           &                $2.62$                   & $$2.92 & $3.92$                            \\ \hline
                                $\frac{\delta \xi^{t}_t}{\xi^{t}_t}$           &  \color{blue}{$-67\%$}              & $210\%$          & $+0.63$           &                $3.60$   & $3.17$ & $3.12$            \\ \hline
  \end{tabular}
\end{center}
As expected, the equity factor $\delta Z_t$ exhibits significant negative skewness and an excess kurtosis, found to be around $1.5$ and inline with numerous previous studies (see~\cite{bouchaud-potters-book}). The small amount of samples should however make us feel suspicious of any definite conclusion. Although there is enough data to prove the existence of significant excess kurtosis and fat tail, there is certainly not enough to calibrate with confidence the corresponding values of $\kappa$. More to the point, estimated tail coefficients in the $3-4$ range raise clearly the question of convergence and of the finiteness of the kurtosis. We ignore this potential issue, and only conclude from the above numbers that the gaussian assumption is clearly violated. \\\\
The volatility is positively skewed, with a much larger kurtosis (with the fast factor being more extreme than the slow factor). This does not come as a surprise as volatility's behaviour is notoriously wild. The tail coefficients, computed from extreme values at the $2\%$ (below) and $98\%$ (above) quantiles, are also representative of the underlying distributions. Positive skew $\zeta\approx +0.5$, significant excess kurtosis $\kappa>3$, and small positive tail coefficient $\nu^+ \approx3$ clearly indicates that implied volatility can behave capriciously, even more as the maturity decreases. 
%
%
\subsubsection{Beware the volatility carry}
Those statistics also highlight two of the most common strategies in the volatility space: the short volatility carry and the implied term-structure carry~\cite{anti-book}. Both strategies consist in selling volatility. They aim at collecting small but regular positive returns by observing a dangerous short volatility position, thereby playing against the rare but devastating risk event of implied volatility spiking.
\begin{itemize}
\item {\bf Short Volatility Carry}  The short volatility aims at harvesting the well known volatility risk premium, i.e. the difference between implied and realized volatilities, by going short realized volatility against the long implied premium. On the current dataset, the volatility risk premium can be estimated to be around $1-\sigma_Z^2 \approx 36\%$ (obviously excluding transaction costs). This is quite large, and explains why the volatility risk premium is so popular. However, the volatility of the risk premium can be roughly approximated around $\sqrt{2+\kappa}\ \sigma_Z^2\approx 120\%$, by no-means insignificant.
\item {\bf Implied Term-Structure Carry} The implied term-structure strategy exploits the negative carry present in the volatility term-structure. The instantaneous variance curve is usually in contango, reflecting the greater uncertainty associated with further maturities. On this dataset, the term-structure premium is characterised by the negative trends of volatility factors, with magnitude above $50\%$! The iPath S\&P 500 VIX Short-Term Futures ETN, which systematically rolls a long position in short-term VIX futures, constitutes probably the most archetypical example of the cost of carry. 
\end{itemize}
Although the above statistics have been computed over the whole period, it should be clear that they are not constant through time. For instance, the ratio of realised over implied volatilities averages at a value around $\sigma_Z \approx 80\%$, but has reached over the considered period very large values. The distribution of the square spot factor $\frac{1}{\delta t}\delta Z_t^2$ is representative of the danger of the short volatility position.\\\\ 
%
Predicting the evolution of a volatility risk premia is difficult. In practice, very few practitioners have been successful over the long run. Yet, despite these obvious dangers, a host of volatility strategies, implementing sophisticated rule-based strategies which are supposedly able to anticipate risk reversals, have recently surfaced and received surprisingly popularity... until the next crisis?
%
%
%
\subsection{Joint Densities and Non-linearities}
\label{jdf}
We now study the characteristics of the joint variations. In addition to the negative correlation between spot and volatility, we expect non-linearities to be present. Volatility tends to react linearly to shocks up to a certain point after which (i.e. below which) volatility tends to spike rapidly.\\\\
We define the centered and normalized variables \[
\delta \bar{Z}_t = \frac{\delta Z_t - \mu_Z \delta t}{\sigma_Z \sqrt{\delta t}} \ \text{and} \ \delta \bar{W}^\alpha_t = \frac{\delta W^\alpha_t - \mu_W^\alpha \delta t}{\sigma_W^\alpha \sqrt{\delta t}},
\]
and display their joint variations in Figure~\ref{fig:jdf}. As we can observe, there exists a small non-linear relationship that is more pronounced for the fast factor.\\
\begin{minipage}{\columnwidth}
\vspace{0.5cm}
\includegraphics[width=\columnwidth]{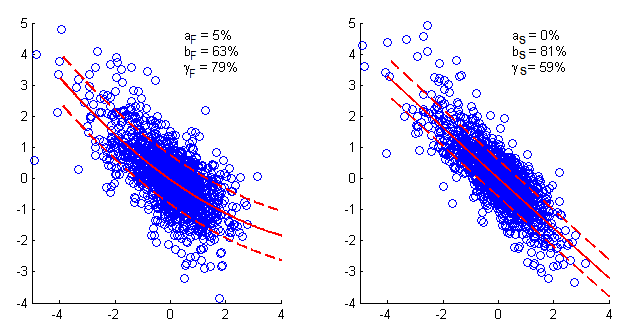}
\captionof{figure}{\it {\bf Joint Dynamics} The fast (left) and slow (right) volatility factors $\delta \bar{W}_t^\alpha$ are plotted against the spot factor $\delta \bar{Z}_t$. We also graph the result of non-linear modeling through the functional relationship $f_\alpha$. The estimated parameters are provided.}
\label{fig:jdf}
\vspace{0.5cm}
\end{minipage}
We suggest to model the daily variations of volatility as the sum of two terms, a non-linear dependancy on the spot factor $\delta Z_t$, and an exogeneous factor modeled by a Gaussian $U_t^\alpha$:
\begin{eqnarray}
\delta \bar{W}_t^\alpha = f_\alpha(\delta \bar{Z}_t) + \gamma_\alpha U_t^\alpha,
\label{eq:nonlinear}
\end{eqnarray}
where the function $f_\alpha$ is chosen as a quadratic function \[
f_\alpha(\delta \bar{Z}_t) = a_\alpha (\delta \bar{Z}_t^2 -1) - b_\alpha \delta \bar{Z}_t.
\]
More complex relationship could have been introduced, but we found this simple quadratic functional to capture well the non-linear dependancy of volatility on the spot. It is interesting to observe that by working at a finite time scale, non-linearities can be easily introduced. This would not be the case with an infinitesimal modeling, as the quadratic variations $d\bar{Z_t}^2=1$ would reduce the functional $f_\alpha$ to a simple linear dependancy.\\\\
Straightforward computations leads to:
\begin{eqnarray*}
a_\alpha &=& \frac{E[\delta \bar{W}_t^\alpha (\delta \bar{Z}_t^2-\zeta \delta \bar{Z}_t)]}{2 + \kappa - \zeta^2},\\
b_\alpha &=& \frac{\zeta E[\delta \bar{W}_t^\alpha \delta \bar{Z}_t^2] -(2+\kappa)E[\delta \bar{W}_t^\alpha \delta \bar{Z}_t]}{2+\kappa-\zeta^2}\approx-E[\delta \bar{W}_t^\alpha \delta \bar{Z}_t],\\
\gamma_\alpha^2 &=& 1 -a_\alpha^2 (2+\kappa) - b_\alpha^2 + 2a_\alpha b_\alpha  \zeta,\\
\rho &=& \gamma_\alpha \gamma_\beta E[U_t^\alpha U_t^\beta] + (2+\kappa) a_\alpha  a_\beta \\
&& +  b_\alpha  b_\beta - \zeta a_\alpha b_\beta - \zeta a_\beta b_\alpha
\end{eqnarray*}
The exogenous variables $U^\alpha_t$ are not strongly correlated.  
Besides, visual inspection seems to indicate independance from the spot variable $\delta \bar{Z}_t$ - note that we have $E[U^\alpha_t \delta \bar{Z}_t] = 0$ by construction. The distance correlation~\cite{szekely-07} between $\delta \bar{W}_t^\alpha$ and $\delta \bar{Z}_t$ is around $70\%$, but falls below $10\%$ between $U^\alpha_t$ and $\delta \bar{Z}_t$.\\\\
Of the two volatility factors $\delta W^\alpha_t$, only the fast one requires a quadratic term. The contribution of the exogenous factor, as measured by $\gamma_\alpha$, is significant in both cases, with a larger impact for the fast factor $\gamma_F > \gamma_S$. Short-term volatility is more wild and less predicatable than longer-term volatility. The use of a third and faster factor, i.e. $k>k_F$, would have generated a slightly higher convexity, but not by significant amount. \\\\
Going one-step further, we deduct from Equation~\ref{eq:var-model} that the variance curve deformations can also be modeled as the sum of two independant terms, one quadratic functional of the spot factor $\delta Z_t$, and an additional independant component:
\begin{eqnarray}
\label{eq:quadravol}
\frac{\delta \xi_t^u}{\xi_t^u} &=& \sum_\alpha{ \theta_\alpha \omega^\alpha(u,t,\xi_t) \left(\mu_\alpha \delta t + \sqrt{\delta t} f_\alpha(\delta \bar{Z}_t)\right)}\\&&+\sigma_V(u,t,\xi_t) \sqrt{\delta t} V_t \nonumber
\end{eqnarray}
where $V_t$ is a standard normal variable and $\sigma_V(u,t,\xi_t)$ a maturity-dependant volatility:
\[
\sigma_V(u,t,\xi_t) = \sqrt{\sum_{\alpha,\beta}{\theta_\alpha\theta_\beta \omega^\alpha(u,t,\xi_t) \omega^\beta(u,t,\xi_t)\gamma_\alpha \gamma_\beta E[U^\alpha U^\beta] }}
\]
The magnitude of the volatility $\sigma_V$ should draw attention to the rather unpredictable nature of implied volatility. The exogenous factor contributes to approximately a third of the total volatility. Although it is certainly true that the dynamics of volatility is strongly linked to its underlying, reducing it to a simple functional relationship would greatly underestimate the subtle behavior of volatility. This naturally leads us to discuss some limitations of GARCH models.\\
\begin{minipage}{\columnwidth}
\includegraphics[width=\columnwidth]{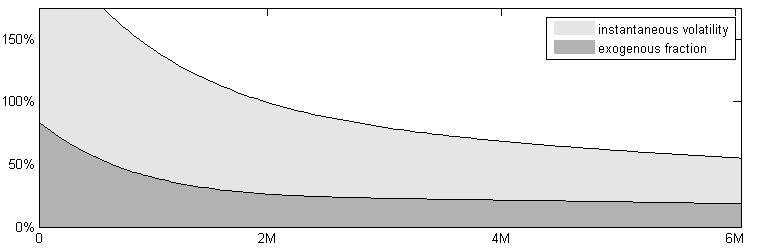}
\end{minipage}
\subsubsection{Link with GARCH models}
\label{garch}
GARCH models are very popular in financial econometrics. They generally assumes that the conditional volatility $\sigma_{G,t}^2 = E_t^\mathfrak{R}[\sigma^2_t]$ is only driven by a single source of risk present in the spot variations $\delta Z_t$. Those models do not attempt to model the implied volatility, but focus instead on the ``true'' variance of future spot returns. As such, they voluntarily ignore the independant nature of volatility. Many variants exist~\cite{engle-2008}, but, in a nutshell, they attempt to model the next step variance $\sigma_{G,t+1}^2$ as a function of past variances $\sigma_{G,t-i}^2$ and past returns $r_{t-i}$ for $i \geq 0$. \\\\
A typical assymetric GARCH$(1,1)$ model would be expressed as:
\begin{eqnarray}
\sigma_{G,t+1}^2 = \phi^0 + \phi^1 \frac{r_t^2}{\delta t}+\phi^2 \frac{r_t}{\sqrt{\delta t}}+(1+\phi^3)\sigma_{G,t}^2+\text{residues}
\label{eq:asmgarch}
\end{eqnarray}
where the constant $\phi^i$ would be calibrated on the historical time series.
The modeling equation~\ref{eq:nonlinear} with its quadratic terms hints at a potential GARCH model. One can develop equation~\ref{eq:quadravol} at first-order to find the following expression:
\begin{eqnarray}
\xi_{t+\delta}^{t+\delta} \approx \phi^0_t + \phi^1_t  \frac{r_t^2}{\delta t} - \phi_t^2 \sqrt{\xi_t^t} \frac{r_t}{\sqrt{\delta t}} +  (1+\phi^3_t)  \xi_t^t  + \xi_t^t \phi^4_t V_t  \sqrt{\delta t}
\label{eq:garch}
\end{eqnarray}
with $V_t$ a standard Gaussian variable and parameters verifying:
\begin{eqnarray*}
\begin{array}{rcl}
\phi^0_t	&	=	&	\frac{\partial \xi_{t}^{u}}{\partial u}|_t\delta t       \\
\phi^1_t	&	=	&	\sum_\alpha{ \bar{\theta}_\alpha\frac{ a_\alpha}{\sigma_Z^2 }} \sqrt{\delta t} \ \text{where} \ \bar{\theta}_\alpha = \theta_\alpha e^{-k_\alpha \delta t} \\
\phi^2_t	&	=	& \sum_\alpha{\bar{\theta}_\alpha  (2 a_\alpha \frac{\mu_Z}{\sigma_Z^2} \sqrt{\delta t} + \frac{b_\alpha}{\sigma_Z} )}\sqrt{\delta t}   \\
\phi^3_t	&	=	&	\sum_\alpha{\bar{\theta}_\alpha \mu_W^\alpha\delta t } - \sum_\alpha{\bar{\theta}_\alpha  a_\alpha \sqrt{\delta t}   } \\
& & + \sum_\alpha{\bar{\theta}_\alpha   (a_\alpha\frac{\mu_Z^2}{\sigma_Z^2}\sqrt{\delta t}+ b_\alpha\frac{\mu_z}{\sigma_Z}) }\delta t  \\
\phi^4_t	&	=	&	\sigma_V(t, t+\delta t,\xi_t)
 	 	 	 	\end{array}
\end{eqnarray*}
The difference between Eq.~\ref{eq:asmgarch} and~.\ref{eq:garch} comes from which variance is modeled; in the former case, it is a measure of the true realized variance $E_t^\mathfrak{R}[\sigma^2_t]$, whereas in the later it corresponds to the implied market view $\xi_t^t=E_t^\mathfrak{M}[\sigma^2_t]$. However, as we can expect market expectations to provide a reasonable estimate of true hidden distribution, both approaches should be equivalent at first-order, with the discrepancy being taken into account in Eq.~\ref{eq:garch} by the presence of the normalizing variance factor $\sigma_Z^2$. \\\\
Calibration of Eq.~\ref{eq:asmgarch} on the current data set shows a good alignment with model values (where we have averaged over the time-dependancy):
\begin{center}
  \begin{tabular}{| c | c | c | c | c | c |}
    \hline
    & $\phi^0$ & $\phi^1$ & $\phi^2$ & $\phi^3$ & $\phi^4$  \\ \hline
    \text{Data} & $0.15\%$ & $1.53\%$ & $0.15\%$ & $-3.13\%$ & $127\%$  \\ \hline
    \text{Model} & $0.00\%$ & $0.96\%$ & $0.15\%$ & $-1.50\%$ & $141\%$ \\
    \hline
  \end{tabular}
\end{center}
\noindent The quadratic term $\phi^1$ is larger when directly calibrated on the data set than computed from model parameters. This is partly due to the implied leverage and the volatility clustering effects that we analyse in the next section. In stochastic model with a term-structure of volatility, a shock $\delta Z_t$ at time $t$ has an impact on the entire variance curve. This impact is reflected in the slope of the term-structure, and indirectly in the coefficient $\phi^0_t$. On the other end, the modeling in Eq.~\ref{eq:asmgarch} and the approximation provided by Eq.~\ref{eq:garch} are near-sighted in the sense that they ignore longer-term past contributions and attempt to explain variance changes through past returns only.\\\\
Stochastic volatility model with term-structure of implied variance, such as the current model of Eq.~\ref{eq:var-model}, are a good imitation of market realities. In particular, long range phenomena are more easily captured thanks to the modeling of term-structure (see sect.~\ref{StylizedFacts}), than for GARCH models\footnote{Long-term dependancies can obviously be introduced in GARCH models, but calibration can then prove difficult as the number of degrees of freedom increases.}. 
\subsubsection{Stylized Facts}
\label{StylizedFacts}
Despite being multifaceted, spot and volatility exhibit a few stable features, which are often referred to as `stylized facts' in the literature; among those,  the negative steepness of the implied volatility smile (reflecting the negative spot/vol correlation), the heterocedasticity of volatility (i.e. the phenomenon of volatility clustering~\cite{cont-01}), and the implied leverage effect (i.e. the tendency of ATM volatility to increase as the underlying market goes down~\cite{ciliberti-bouchaud-potters-leverage-08}).\\\\
Having modeled the non-linear spot/vol relationship, we study the modeling implication on the implied leverage and the volatility clustering effects. Both relates today's news to tomorrow's volatility. The implied leverage effect quantifies the impact of a shock today onto tomorrow's volatility (sign to amplitude), whereas the volatility clustering relates today's volatility to tomorrow's (amplitude to amplitude). In our model, the straightforward link is obviously provided by the term-structure of volatility: a news today has an impact on the full term-structure of variance, thereby impacting tomorrow's volatility.\\\\
We denote by $\widehat{X}$ the detrending of the stochastic variable $X$, i.e. $\widehat{X}=X-E[X]$. The leverage correlation function $\mathcal{L}(t,\Delta)$ measures the dependancy between a (detrended) return $\widehat{r}_t$ observed today at time $t$ and tomorrow's volatility $\widehat{r_{t+\Delta}^2}$ measured at time $t+\Delta$. Straightforward computation by iterative conditioning on the filtrations at times $t+\Delta$ and $t+\delta t$, followed by the use of Eq.~\ref{eq:var-model} (see~\cite{vargas-2013}) leads to:
\begin{eqnarray*}
\mathcal{L}(t,\Delta) &=& \frac{E_t[\widehat{r}_t \widehat{r_{t+\Delta}^2}]}{\sqrt{E_t[\widehat{r}_t^2]}E_t[\widehat{r_{t+\Delta}^2}]}\\
&=&\sum_{\alpha}{\theta_\alpha \omega^{\alpha}(t,t+\Delta,\xi_t) \underbrace{E[\delta \bar{Z}_t \delta \bar{W}^{\alpha}_{t}]}_{a_\alpha \zeta - b_{\alpha}\approx -b_\alpha}} \sqrt{\delta t}
\end{eqnarray*}
Therefore, in the case of equities, the main driver of the leverage correlation function is the spot/vol correlation $E[\delta \bar{Z}_t \delta \bar{W}^{\alpha}_{t}] \approx -b_\alpha$. Non-linearities are negligeable.\\\\
The volatility clustering function $\mathcal{C}(t,\Delta)$ measures the correlation between today's volatility $\widehat{r_{t}^2}$ computed at time $t$ and tomorrow's volatility $\widehat{r_{t+\Delta}^2}$ computed at time $t+\Delta$. Using similar derivation steps, we find that:
\begin{eqnarray*}
\mathcal{C}(t,\Delta) &=&\frac{E_t[\widehat{r_t^2} \widehat{r_{t+\Delta}^2}]}{E_t[\widehat{r_t^2}] E_t[\widehat{r_{t+\Delta}^2}]}]\\
&=&\sum_{\alpha}{\theta_\alpha\omega^{\alpha}(t,t+\Delta,\xi_t)  \underbrace{E[\delta \bar{Z}_t^2 \delta \bar{W}^{\alpha}_{t}]}_{a_\alpha(2+\kappa)-b_\alpha \zeta}}\sqrt{\delta t}
\end{eqnarray*}
In the case of volatility clustering, the complete non-linear relationships between the spot returns and the volatility factors must be taken into account. It is interesting to note that if we had assumed a standard normal modeling, i.e. assuming the spot factor $\delta Z_t$ to be Gaussian by enforcing $\zeta = 0$ and $\kappa=0$, we would not have been able to match the data as accurately. The volatility clustering is as much a result of non-linearities (contributing a third, through the term $a_\alpha (2+\kappa)$) as of non-Gaussian effects (contributing two thirds, through the term $b_\alpha \zeta$) .\\
{
\begin{minipage}{\columnwidth}
\includegraphics[width=0.95\columnwidth]{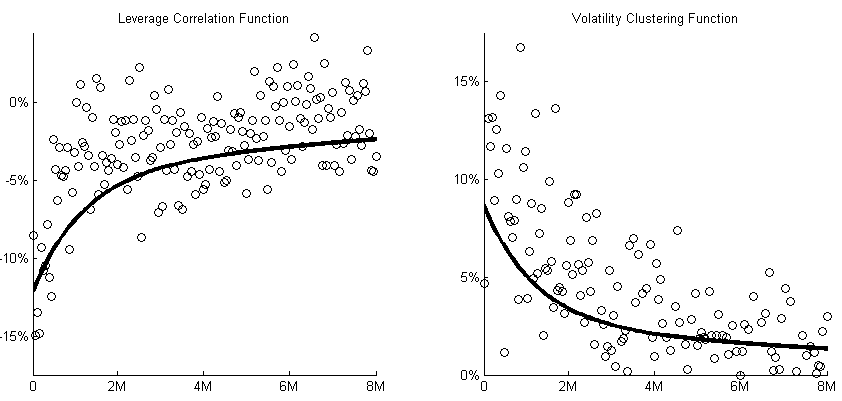}
\captionof{figure} {\it {\bf Leverage Correlation $\&$ volatility Clustering}}
\label{fig:lcvc}
\vspace{0.5cm}
\end{minipage}
}
Figure~\ref{fig:lcvc} displays the leverage correlation function and the volatility clustering function estimated from our model and the data set. The augmented stochastic model captures well both stylized facts.\\
%
%
%
%
\subsection{Towards a term-structure of vol of vol}
\label{VoV}
Although spot returns $r_t=\frac{\delta S_t}{S_t}$ display little autocorrelation at the scale of a day\footnote{A small negative auto-correlation seems to exist for some equity indices, giving rise to mean-reversion strategies.}, this is not the case of the return magnitudes, measured as $|r_t|$ or $r_t^2$. This phenomenon is known as the volatility clustering effect that we studied in details in the previous section. \\\\
In our model, the variables $\delta Z_t$ and $\delta W^\alpha_t$ are assumed to be independant and identically distributed. As such, they should not show any autocorrelation. This is verified by the spot factors $\delta Z_t$, their magnitudes $|\delta Z_t|$, as well as by the volatility factors $\delta W_t^\alpha$. 
However, the magnitudes of volatility factors $|\delta W_t^\alpha|$ are not independant and a positive autocorrelation exists. A simple autocorrelation check over the whole term-structure confirms the finding: the absolute variations of the curve observed at different maturities are also autocorrelated (a similar check can be conducted on the variations of VIX futures).\\\\
This autocorrelation property hints at a similar behaviour for the variations of volatility $\frac{\delta \xi_t^u}{\xi_t^u}$ as for the spot returns $r_t$ : in the same way that a shock today has a lasting impact on the future market volatility, a volatility shock today also impacts the future volatility of volatility. As the former translates into a term-structure of implied-volatility, a similar term-structure of volatility of volatility also exists. \\\\
To better understand the behaviour of volatility of volatility, we look into the VVIX index. Similarly to the VIX index, which represents the expected volatility  of the SPX index over the next 30 calendar days, the VVIX index reflects the expected volatility of the VIX index. In the same vein, it is computed as an interpolation of VIX options, which are written on VIX futures. Because VIX futures react differently to the arrival of new information based on their time to expiry, the situation is slightly different than for the VIX index. The volatility of short-term VIX futures is naturally higher than the volatility of further-away VIX futures, thereby generating a term-structure that is reflected in the computation of the VVIX index. This is not the case with the VIX index that is computed from options that are all written on the same underlying.\\\\
In our model with constant vol of vol parameters, and assuming a flat term-structure of volatility, the expected total variance of a VIX future of maturity $T_i$ can be evaluated to be: 
\begin{eqnarray}
E_t[\frac{1}{T_i-t}\int_t^{T_i}{(\frac{d \mathcal{V}_u^{T_i}}{\mathcal{V}_u^{T_i}}})^2] = \frac{1}{4}\sum_{\alpha,\beta}{\Omega^{\ g}_{\alpha,\beta}\  \underbrace{g((k_\alpha+k_\beta)(T_i-t))}_{\text{denoted }g_{\alpha+\beta}(T_i-t)} }
\label{eq:vvixts}
\end{eqnarray}
with $\Omega^{\ g}_{\alpha,\beta} = \Omega_{\alpha,\beta}g_\alpha(\Delta T) g_\beta(\Delta T)$. The term-structure of the volatility of VIX futures is therefore decreasing - as we just mentionned, further-away VIX futures react less to news, and as such, are less volatile. A comparison with historical data (the VVIX term-structure is made available by the Chicago Board Options Exchange) shows that our model under-estimates significantly the volatility of volatility embedded in the pricing of VIX options. This is expected. Similarly to the volatility risk premium, a volatility of volatility risk premium exists. Careful inspection also reveals that the term-structure is dependant on the level of vol of vol : with higher vol of vol (i.e. higher VVIX), the historical term-structure becomes steeper. This clearly shows a limitation of our approach: with our constant vol of vol parameters $\Omega$, the term-structure of Eq.~\ref{eq:vvixts} is decreasing but does not change through time.\\\\
Going one step further, we compute the model value of the VVIX index in our model. Because of the decreasing term-structure and the interpolation methodology, the value depends on the maturity of the chosen expiries (again, this is not the case with the VIX index). In our model, the square of the expected VVIX index should be equal to:
\[
\sum_{\alpha,\beta}{\frac{\Omega_{\alpha,\beta}}{4} g_\alpha g_\beta \left ( 2 g_{\alpha+\beta}(T_2-t) - e^{-(k_\alpha+k_\beta)(T_1-t)} g_{\alpha+\beta}(\Delta T)\right ) }
\]
with $T_1$ and $T_2$ the first and second expiries used in the interpolation process and $\Delta T = T_2-T_1\approx \frac{30}{365}$. This implies an average VVIX value at around $75\%$ that is significantly lower than the realized historical average (at around $86\%$), thereby implying that the vol of vol risk premium embedded in VIX options is significative.\\\\
Figure~\ref{fig:vvix} represents the ratio between the squared official VVIX index and our model. The ratio that we denote $\lambda_t$ appears to be mean-reverting towards a value that can be estimate at around $130\% \approx (\frac{86\%}{75\%})^2$. \\
\begin{minipage}{\columnwidth}
\includegraphics[width=\columnwidth]{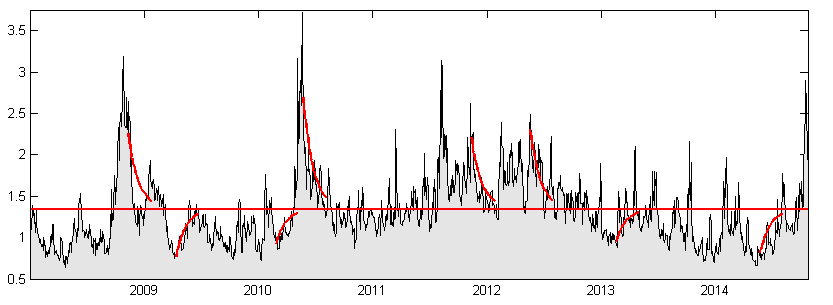}
\captionof{figure}{\it {\bf Volatility of Volatility} We display the ratio between the square VVIX index and our theoritical value, computed with fixed vol of vol parameters. }
\label{fig:vvix}
\vspace{0.5cm}
\end{minipage}
The dynamics of the ratio suggests that a mean-reverting process could be used to model the vol of vol randomness. For instance:
\[
\frac{d \lambda_t}{\lambda_t} = - k_\lambda (\log{\lambda_t} - (\log{\lambda_\infty}+\frac{\sigma_\lambda^2}{2k_\lambda}))dt + \sigma_\lambda dW^\lambda_t
\]
with calibrated parameters 
\begin{center}
\begin{tabular}{| c | c | c | c | c | c |}
\hline
$\lambda_\infty$ & $k_\lambda$ & $\mu_\lambda$ & $\sigma_\lambda$ & $\zeta_\lambda$   & $\kappa_\lambda$   \\ \hline
$126\%$ &  $16$ & $0\%$ & $152\%$ & $0.78$   & $2.66$  \\ \hline
\end{tabular}
\end{center}
and correlation with other factors
\begin{center}
\begin{tabular}{| c | c | c | c |}
\hline
$\rho$ & $\delta Z_t$ & $\delta W^F_t$   & $\delta W^S_t$   \\ \hline
$\delta W^\lambda_t$ & $-60\%$ & $56\%$ & $59\%$ \\\hline
\end{tabular}
\end{center}
This would generate a term-structure of vol of vol equal to
\begin{eqnarray}
\label{vovts}
E_t[\log{\lambda_u}]=\log{\lambda_\infty}+(\log{\lambda_t}-\log{\lambda_\infty})e^{-k_\lambda(u-t)}
\end{eqnarray}
Within this augmented framework, the instantaneous variance would follow a diffusion equation:
\begin{eqnarray*}
\frac{d \xi_t^u}{\xi_t^u} &=& \sqrt{\frac{\lambda_t}{\lambda_\infty}}\times \sum_{\alpha=1}^{n} {\theta_\alpha \omega^{\alpha}(t,u,\xi_t) d W^{\alpha}_t}
\end{eqnarray*}
At first-order in the vol of vol parameters, only minor adjustments to our stochastic volatility framework would be required - however, it would become less tractable.
%
The convexity correction would be altered, 
so that with higher vol of vol, the convexity correction would be slightly increased. 
More importantly, the presence of stochastic vol of vol means that the total variance of a VIX future of maturity $T_i$ would reflect the mean-reverting behaviour of vol of vol. By using a second-order approximation derived from Eq.~\ref{vovts}, 
\[
\lambda_t^u = E_t[\lambda_u] = \lambda_\infty (\frac{\lambda_t}{\lambda_\infty})^{e^{-k_\lambda(u-t)}}e^{\frac{\sigma_\lambda^2 (u-t)}{2}},
\]
the total variance of a VIX future of maturity $T_i$ would differ from eq.~\ref{eq:vvixts}:
\begin{eqnarray*}
\sum_{\alpha,\beta}{\frac{\Omega^{\ g}_{\alpha,\beta} e^{-\frac{\sigma_\lambda^2}{2}t  -(k_\alpha+k_\beta)T_i}}{4(T_i-t)}\int_t^{T_i}{\frac{\lambda_t^u}{\lambda_\infty}e^{(k_\alpha+k_\beta+\frac{\sigma_\lambda^2}{2})u }  du}}
\end{eqnarray*}
The impact of vol of vol would be minimal for short-maturities, small mean-reversion rate $k_\lambda$, or small level of vol of vol. In all other cases, the above integral would need to be evaluated numerically. As the vol of vol increases, the term-structure becomes steeper as expected from historical data. We keep for future work the integration of vol of vol information in the model.
%
%
%
%
%
%
%
%
%
%
%
\section{Consequences}
\label{consequences}
We pursue our exploration of the spot/vol properties by delving into the theoritical implications of the non-linearities. We investigate the impact on the pricing and hedging of derivatives, an area of active research~\cite{backus-1997,bergomi-smiledynamicsIV,ciliberti-bouchaud-potters-leverage-08,bergomi-guyon-2012,vargas-2013}. We first focus on standard vanilla options and review the link between skewness of an underlying and the implied skew (section~\ref{skewnessskew}). We then scrutinize the relationship between implied skew and the evolution of the ATM implied volatility (section~\ref{SSR}). Finally, we look at some of the implications on the volatility of annualized variance (section~\ref{VVS}).
\subsection{Skewness and Skew}
\label{skewnessskew}
We investigate the relationship between skewness of returns $\zeta_t^T$ and the skew  $\text{Skew}_{t,T}$ of implied option smile.
It is a well-known fact that the skewness of the returns generated by a model is linked to the implied skew of the options priced by the same model~\cite{backus-1997}. This is not surprising, since the skewness and the skew are both functions of the spot-volatility correlation. At first-order in vol of vol parameters, the relationship can be expressed as $\text{Skew}_{t,T} = \frac{\zeta_t^T}{6\sqrt{T-t}}$. However, this expression is only true when the model is linear. When some non-linearities are present, e.g. through the function $f_\alpha$, the equality breaks-down as pointed out in~\cite{vargas-2013}.\\\\
We investigate the relationship within the limits of our model. We verify that when non-linearities are ignored, i.e. assuming $a_\alpha=0$, the equality is valid at first-order. The presence of non-linearities alters this relationship. However, in the case of equities where the linear spot/vol correlation dominates, the impact is negligeable.
\subsubsection{Skewness of returns}
The skewness of the returns can easily be computed from of the moments of order $2$ and $3$:
\[
\zeta_t^T = \frac{E_t[(\widehat{\log{\frac{S_T}{S_t}}})^3]}{E_t[(\widehat{\log{\frac{S_T}{S_t}}})^2]^\frac{3}{2}}.
\]
Following the same derivation steps as in~\cite{bergomi-smiledynamicsII}, we find at first-order that the skewness can be expressed as:
\begin{eqnarray}
\zeta_t^T \approx \frac{\sum_u{(\xi_t^u \delta u)^\frac{3}{2}  }}{(\sum_u{\xi_t^u\delta u})^\frac{3}{2}} \zeta_Z \hspace{4.5cm}\\
+  3\sum_{\alpha}{\theta_\alpha\frac{\sum_{u}{\xi_{t}^u \sum_{v<u}{\sqrt{\xi_t^v} \omega^{\alpha}(v,u,\xi_t) E[\delta \bar{Z}_u d \bar{W}^{\alpha}_{u}]\delta u \delta v}}}{(\sum_u{\xi_t^u\delta u})^\frac{3}{2}}} \nonumber
\end{eqnarray}
Assuming a relatively flat-term structure of volatility, the above equation further simplifies to
\begin{eqnarray*}
\zeta_t^T \approx  \frac{\zeta}{\sqrt{N}} + 3\sqrt{T-t}\sum_{\alpha}{\theta_\alpha \underbrace{E[\delta \bar{Z} d \bar{W}^{\alpha}]}_{-b_\alpha}h(k_\alpha(T-t))}
\end{eqnarray*}
where $h(x)$ is defined as $h(x) = \frac{1}{x^2}\int_0^x{ug(u)du}=\frac{x-1+e^{-x}}{x^2}$.
The skewness of the returns for maturity $T-t$ is the result of the intrinsic skewness of the spot process at time scale $\delta t$, and of the spot-volatility correlation. With no vol of vol, i.e. $\theta_\alpha=0$, the skewness is decreasing in $\frac{1}{\sqrt{N}}$ as expected for a process with independant increments. This term quickly becomes negligeable.
\subsubsection{Smile of volatility}
To investigate the impact of vol of vol on the implied smile of options, we introduce a scaling parameter $\lambda$ as $\theta_\alpha \to \lambda \theta_\alpha$. The parameter $\lambda$ controls the amount of stochastic volatility in the model. With no vol of vol, i.e. $\lambda=0$, our model has constant volatility equal to the var-swap volatility $\sigma_{\text{VS}}(t,T) = \sqrt{\mathbb{V}_{t}^{t\to T}}$, and the implied volatility smile of options is flat. In the presence of vol of vol, the shape of the implied volatility surface is altered, the ATM volatility shifts and the skew deviates from zero. This property is a well-known fact of stochastic volatility models, and has been quantified accurately in the case of linear model at second-order in recent work~\cite{bergomi-smiledynamicsIV,bergomi-guyon-2012}.\\\\
When non-linearities are present, the impact of stochastic parameters on the implied volatility is different, as pointed out in~\cite{vargas-2013} in the case of a single-factor GARCH model. We conduct a similar analysis and compute the implied volatility smile in our augmented non-linear stochastic volatility model. 
At first-order in the strike $K=S_t+dK$, the volatility smile is approximated by:
\[
\sigma^\lambda(K,t,T) \approx \sigma^\lambda_{\text{ATM}}(S_t,t,T) +  \text{Skew}^\lambda_{t,T} \times \frac{dK}{S_t}
\]
where $\sigma^\lambda(K,t,T)$ is the Black-Scholes implied volatility observed at strike $K$. Note that $\sigma^{\lambda=0}(K,t,T)=\sigma_{\text{VS}}(t,T)$ for all strikes $K$.
The price of a call option $F_K(\lambda) = E_t[(S_T-K)^+]$ of strike $K$ is function of the vol of vol parameters through the parameter $\lambda$. Pricing is achieved under the risk-neutral measure, i.e. we assume that $\delta \bar{Z}_t$ and $\delta \bar{W}_t$ are standard normal distributions, and we neglect the drift component. At first-order, the volatility shift implied by the presence of stochastic volatility at a specific strike $K$ can be computed as:
\begin{eqnarray*}
\delta \sigma (K,t,T) = \sigma^\lambda(K,t,T)-\sigma_{\text{VS}}(t,T) =  \lambda \frac{F'_K(0)}{\text{Vega}_K}
\end{eqnarray*}
where the vega is the standard Black-Scholes vega. We deduct that:
\begin{itemize}
\item {\bf ATM Spread} The ATM volatility is shifted by $\sigma^\lambda(S_t,t,T)-\sigma_{\text{VS}}(t,T) = \lambda \frac{F'_{S_t}(0)}{\text{Vega}_{S_t}}$
\item {\bf Skew} The skew generated by the stochastic volatility is equal to
\[
\text{Skew}^\lambda_{t,T} = \lambda \left(\frac{F'_{K}(0)}{\text{Vega}_{K}}-\frac{F'_{S_t}(0)}{\text{Vega}_{S_t}}\right) \frac{S_t}{dK}
\]
\end{itemize}
After some tedious computations (see app.~\ref{app:smile}), we find that the ratio $ \frac{F'_K(0)}{\text{Vega}_K}$ can be expressed as:
\begin{eqnarray*}
\sum_\alpha {\frac{\theta_\alpha}{2} \frac{ \sum_{u}{\left [\xi_t^u \delta u \sum_{v<u}{\delta v \omega^\alpha(u,v) \frac{1}{\sqrt{\delta t}} E_t[ f_\alpha ( A_v+B_v W)]  }\right]}}{(T-t)\sqrt{\mathbb{V}_{t}^{t\to T}}} }
\end{eqnarray*}
where $W$ is a standard Gaussian variables and $A_v, B_v$ are defined as:
\begin{eqnarray*}
A_v &=& \frac{\sqrt{\xi_t^v\delta v}}{(T-t)\mathbb{V}_{t}^{t\to T}}  (\frac{1}{2}(T-t)\mathbb{V}_{t}^{t\to T}+\log\frac{K}{S_t})\\
B_v &=& \sqrt{1-\frac{\xi_t^v\delta v}{(T-t)\mathbb{V}_{t}^{t\to T}}}
\end{eqnarray*}
The expression above is the same as the one derived in~\cite{vargas-2013}. This is not surprising as the exogenous volatility contributions $U_t^\alpha$ have no measurable impact on the expression of the implied smile or skew. From the definition of the function $f_\alpha$, we find that the above expectation can be expressed as:
\begin{eqnarray*}
E_t[ f_\alpha (A_v+B_v W)] &=&  a_\alpha(A_v^2+B_v^2-1) -b_\alpha A_v
\end{eqnarray*}
\underline{\bf  Linear Case} We first consider a linear model and ignore quadratic terms by simply assuming $a_\alpha=0$. It is then straightforward to check that the expression $\text{Skew}_{t,T} = \frac{\zeta_t^T}{6\sqrt{T-t}}$ becomes valid at first-order (remember that for pricing we assume that $\zeta=0$ - we neglect the intrinsic skewness of the returns). In addition, one can express exactly the value of the ATM spread and skew implied by the linear model. For the sake of simplicity, we assume that the term-structure of variance is relatively flat; we find that:
\begin{eqnarray*}
\text{Spread}|_\text{lin} &=& -\sum_{\alpha}{\frac{\theta_\alpha b_\alpha}{4}h(k_\alpha (T-t))}(T-t)\sigma^2_{\text{VS}}(t,T)\\
\text{Skew}|_\text{lin} &=& -\sum_{\alpha}{\frac{\theta_\alpha b_\alpha}{2}h(k_\alpha (T-t))}
\end{eqnarray*}
This is exactly the results derived in a more general settings in~\cite{bergomi-guyon-2012}. The presence of vol of vol decreases the ATM volatility and generates skew proportionally with the equality $\text{Spread}|_\text{lin} = \frac{(T-t)\sigma^2_{\text{VS}}(t,T)}{2}\text{Skew}|_\text{lin}$.\\\\
\underline{\bf Non-Linear Case} In the presence of non-linearities, the skew and the skewness are no longer directly related. In the case of flat term-structure, we can express the difference $\text{Skew}_{t,T} - \frac{\zeta_t^T}{6\sqrt{T-t}}$ as:
\[
\sum_{\alpha}{\frac{\theta_\alpha a_\alpha h(k_\alpha (T-t))}{2}}\sigma_{\text{VS}}\sqrt{\delta t}
\]
The impact of non-linearities is measured by the magnitude of the ratio $\frac{a_\alpha }{b_\alpha}\sigma_{\text{VS}}\sqrt{\delta t}\approx \frac{f_\alpha^{''} }{2f_\alpha^{'}}\sigma_{\text{VS}}\sqrt{\delta t}$. In the case of the SPX index, the dominant factor remains the linear spot/vol correlation. Non-linear effects are negligeable and the ratio is close to zero. However, when the correlation becomes small and/or non-linearities large, the difference and the ratio will become observable. This would be the case for assets where the smile is less steep and for which the spot/vol correlation is close to zero, such as FX assets. Note also that when the time frame of observation becomes small, i.e. $\delta t \to 0$, we end up with the linear case. \\\\
One can also compute the impact on the spread of ATM volatilities to find:
\begin{eqnarray*}
\Delta Spread &=& \text{Spread}|_\text{non-lin} - \text{Spread}|_\text{lin}\\
 &=& \sum_{\alpha}{\frac{\theta_\alpha a_\alpha h(k_\alpha (T-t))}{2}}\sigma_{\text{VS}}\sqrt{\delta t}(\frac{(T-t) \sigma^2_{\text{VS}} }{4}-1)
\end{eqnarray*}
which we find to be also negligeable. Although non-linearities would alter the relationship between skewness of the returns and skew of the implied volatility , the magnitude of the correction is small and can safely be ignored.
\subsection{Skew-Stickiness ratio}
\label{SSR}
In recent work~\cite{bergomi-smiledynamicsIV}, Bergomi showed that two a-priori very distinct features of a volatility model, the static shape of the implied smile and the dynamics of the ATM volatility, were strongly linked. To measure their dependancy, he introduces the Skew-Stickiness Ratio $R(t,T)$ as:
\[
R(t,T) = \frac{E_t[\delta \sigma_\text{ATM}(t,T) \delta S_t]}{\text{Skew}(t,T) E_t[dS_t^2]} 
\]
This ratio provides a quantitative interpretation to the notion of sticky-strike $R=1$, sticky-delta $R=0$, and local-vol strike $R=2$.\\\\
%
The correlation between the ATM volatility and the spot variations can be easily computed in our model:
\[
\frac{E_t[\delta \sigma_\text{ATM}(t,T) \delta S_t]}{E_t[\delta S_t^2]} = \sum_\alpha{\frac{\theta_\alpha\int_t^T{ \xi_t^u w^\alpha(t,u)E_t[\delta W^\alpha_t\delta  Z_t]} }{2\sqrt{\xi_t^t \mathbb{V}_{t}^{t\to T}}(T-t)}\delta t}
\]
which in the case of flat term-structure of volatility can be expressed as
\[
\sum_\alpha{\frac{\theta_\alpha}{2}g(k_\alpha (T-t))}E[\delta \bar{Z}_t \delta \bar{W}^\alpha_t]
\]
with $g(x) = \frac{1-e^{-x}}{x}$. 
%
%
%
Ignoring non-linearities (setting $a_\alpha=0$), we therefore find that the skew-stickiness ratio can be expressed as 
\[
R(t,T) = \frac{\sum_\alpha{\theta_\alpha b_\alpha g(k_\alpha (T-t)}}{\sum_\alpha{\theta_\alpha b_\alpha h(k_\alpha (T-t)}}
\]
which is exactly the expression found by Bergomi in~\cite{bergomi-smiledynamicsIV}. In the limit of small maturities, the skew-stickiness ratio converges to 2. This values makes perfect sense if one interprets the smile as the average over all paths of volatilities weighted by the gamma of the option\footnote{
We provide a simple intuitive proof. At time $t$, the volatility smile $\sigma_\text{BS}(K)$ observed for maturity $t+2\delta t$ can be approximated for strikes around the money by $\sigma_\text{BS}(K = S_t+dK) = \sigma_\text{BS}(S_t) + \text{Skew}_t \frac{dK}{S_t}$. At time $t$, the local volatilities corresponding to the time intervals $[t,t+\delta t]$ and $[t+\delta t,t+2\delta t]$ are denoted $\sigma_t$ and $\sigma_{t+\delta t}(S_t+dK)$ respectively. In the limit of small $\delta t$ and small $dK$, we must have $\sigma_\text{BS}^2(K = S_t+dK) = \frac{1}{2}\sigma_t^2 + \frac{1}{2} \sigma_{t+\delta t}^2(S_t+dK)$. The local volatility $\sigma_{t+\delta t}(S_t+dK)$ represents the time-$t$ expectation of the ATM volatility defined on the interval $[t+\delta t,t+2\delta t]$. By writting $\sigma_{t+\delta t}(S_t+dK) = \sigma_{t+\delta t}(S_t) + \text{R}\times \text{Skew}_t \frac{dK}{S_t}$ with $\text{R}$ an unknown coefficient representing the Skew-Stickiness Ratio, we then immediately find that $\text{R} = 2$.
}. 
In the case of long-maturities, the ratio converges towards 1.\\\\
Although non-linearities alter the value of the skew-stickiness ratio, we did not find the differences to be significative for the SPX index.
\subsection{Volatility of Variance Swap}
\label{VVS}
In this final section, we step into the world of volatility derivatives~\cite{carr-lee-volatility-09,ayache-09} and study the volatility of the annualized variance. 
A variance swap provides an exposure to the aggregated annualized variance of the returns during a fixed period $[T_1,T_2]$. Most often, the returns are computed close to close, but other conventions exist. During the life of the trade $T_1\leq t \leq T_2$, the annualised variance mark-to-market
\[
\mathbb{V}_t^{T_1\to T_2}=E_t[\frac{1}{T_2-T_1}\text{var}^{T_1\to T_2} ]
\]
changes according to daily returns that increase the aggregated realised variance, but also due to changes in implied volatility corresponding to the remaining variance up to maturity. \\\\
%
Using the additivity of variance, the variation of the annualised variance $\delta \mathbb{V}_{t}^{T_1\to T_2}$ during time step $\delta t$ (corresponding to a day) can be written as the sum of two explicit terms:
\begin{eqnarray}
\label{eq:vardecompo}
\underbrace{\frac{1}{\Delta T}\left({(\frac{\delta S_t}{S_t} )}^2- \xi^t_t \delta t \right)}_{\text{accrued realised}} + \underbrace{\frac{T_2-t}{\Delta T}\left ( \delta \mathbb{I}_t-E_t[\delta \mathbb{I}_t]\right)}_{\text{variation of implied variance}}
\end{eqnarray}
where $\mathbb{I}_t=\mathbb{V}_t^{t\to T_2}$ denotes the implied annualized variance up to maturity\footnote{The instantaneous implied variance $\xi_t^t$ must verify $\xi_t^t \delta t= \mathbb{I}_t \delta t - (T_2-t)E_t[\delta \mathbb{I}_t]$.}. This expression shows clearly that the total volatility of the annualised variance, i.e. the square-root of
\[
E_{T_1}[\frac{1}{\Delta T}\int_{T_1}^{T_2}{(\frac{\delta \mathbb{V}_u}{\mathbb{V}_u})^2}],
\]
depends on the covariance parameters $\Omega_{\alpha,\beta}$ (through the second term), but also on the kurtosis $\kappa$ of the normalized returns $\delta Z_t$ (through the first-term). This implies that, even under an idealised scenario with no vol of vol, i.e. $\Omega_{\alpha,\beta}=0$,  the discretisation of the returns generates some volatility. This contribution of discrete sampling to the total variance is well-known.  \\\\
A third contribution exists, although it has been less documented. Large unexpected shocks, i.e. $\delta \bar{Z}_t^2 >> 1$, also contributes to the total volatility through their correlation with the implied volatility. This last-term is a direct consequence of non-linearities, with large unexpected shocks, often negative, being stronly correlated with implied variance jumps. We denote by $\rho_{\text{shocks}}^{\alpha}$ this correlation, i.e. $\rho_{\text{shocks}}^{\alpha} = E_t[\delta \bar{W}_t^{\alpha} \times \frac{\delta \bar{Z}^2-1}{\sqrt{2+\kappa}}] $. Using the approximation defined in Eq.~\ref{eq:nonlinear}, we can compute the correlation explicitly to be $\rho_{\text{shocks}}^\alpha = a_\alpha \sqrt{\kappa + 2}-b_\alpha\frac{\zeta}{\sqrt{\kappa + 2}}$. From our estimated parameters, both correlations are of order $25\%$. Similarly to the volatility clustering effect, it is essentially the skewness of the variable $\delta Z_t$ associated to the negative spot-vol correlation that is responsible for the magnitude of the correlation coefficients.\\\\
Under our usual assumption of relatively flat term-structure of variance (see app.~\ref{app:variance} for the derivations), the total variance of a variance swap can be approximated as the sum of three-terms:
\begin{eqnarray*}
\begin{array}{ll}
\text{ sampling impact} & \frac{\kappa + 2}{N\Delta T} \\
\text{ implied parameters} & \sum_{\alpha,\beta}{\Omega_{\alpha,\beta} l_(k_\alpha,k_\beta,\Delta T)}  \\
\text{ unexpected shocks } & 2\sqrt{\frac{\kappa + 2}{N\Delta T}}\sum_{\alpha}{\rho_{\text{shocks}}^{\alpha} \theta_\alpha h(k_\alpha \Delta T)}
\end{array}
\end{eqnarray*}
with the functions defined by $h(x) = \frac{x-1+e^{-x}}{x^2}$ and $l(x,y,z)=\frac{1}{x y z}(\frac{1}{z} - \frac{1-e^{-x z}}{x z^2}-\frac{1-e^{-y z}}{y z^2}+\frac{1-e^{-(x+y) z}}{(x+y)z^2})$. For small maturities, they verify $l(k_\alpha,k_\beta,\Delta T) \approx \frac{1}{3}$ and $h(k_\alpha \Delta T) \approx \frac{1}{2}$, whereas for large maturities, $g_{\alpha,\beta}(\Delta T) \approx \frac{1}{k_\alpha k_\beta \Delta T^2}$ and $h(k_\alpha \Delta T) \approx \frac{1}{k_\alpha \Delta T}$.
Based on the estimated model parameters, we can compute the total variance qualitatively. As Fig.\ref{fig:varvov} illustrates, all three terms have a significant impact, including non-linear effects. In fact, the impact of large unexpected shocks should not be overlooked, as its contribution for short-term variance swaps can be large (e.g. of the order $10\%$ for 3-month swaps). \\
\begin{minipage}{\columnwidth}
\label{fig:varvov}
\vspace{0.5cm}
\includegraphics[width=\columnwidth]{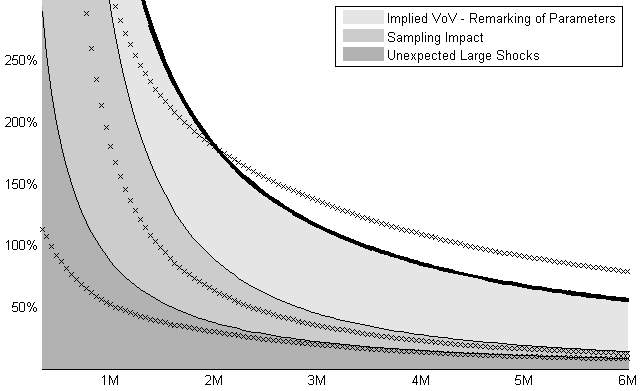}
\noindent {\it {\bf Variance Swap Variance} The expected variances of a variance swap are plotted as a function of maturity. The variance is decomposed into three-terms reflecting the impact of non-linearities, of the discrete sampling, and of the remarking of implied parameters. We also display (black crosses) historical variances computed on our dataset.}
\vspace{0.5cm}
\end{minipage}
The pricing and hedging of volatility derivatives, such as volatility swaps, and options on volatility or variance, depends on the total variance that we have just calculated. As non-linearities contribute to increase the total variance, they would impact the pricing and hedging of derivatives. In this context, the three terms can be directly interpreted as the cost of gammas, the gamma on spot, the gamma on the remarking of parameters, and the cross-gamma between both. However, it is important to realize that the daily hedging of a variance derivative using spot variance swaps of same maturity $T_2$ would hedge the three gammas at once. Said differently, as long as one knows the correct hedging vega, the three terms would be hedged at once. \\\\
As a final comment, we note that the term-structure of variance is rarely flat. In practice, the presence of a slope should be integrated in the total variance (see app.~\ref{app:variance}). As a result, volatility derivatives, such as options on variance, would require additional variance swap hedges of intermediate maturities $T_1<t<T_2$. 
\section{Conclusion}
\label{conclusion}
In this paper, we investigated some characteristics of spot and volatility from an empirical perspective. We summarise below our findings:
\begin{enumerate}
\item A Karhunen-Lo\`eve decomposition of the variance curve deformations shows that the first two eigenmodes account for almost $99\%$ of the variance, corroborating results reported in~\cite{cont-fonseca-01}. A stochastic volatility model with only two factors is able to capture with a high degree of accuracy the daily variations of the variance curve up to half a year.
\item The densities of the spot and volatility factors deviate markedly from normal well-behaved distributions. They exhibit significant skew, large excess kurtosis and fat tails, an known fact that has been frequently documented~\cite{bouchaud-potters-book}. The relationship between spot and volatility is not linear; volatility is convex. In the case of the SPX index, accurate modeling can be achieved with a quadratic functional $f_\alpha$. As maturity increases, the magnitude of the non-linear component weakens. The fraction of the variance of variance unexplained by the functional relationship amounts to a third.
\item The leverage correlation, which quantifies the correlation between a spot move today and tomorrow's realised volatility, is accurately modeled by the term-structure of implied variance (Eq.~\ref{eq:var-model}) and by the linear correlation between spot and volatility~\cite{ciliberti-bouchaud-potters-leverage-08,vargas-2013}. 
\item The modeling of the volatility clustering, which measures the correlation between today's and tomorrow's realised volatilities, requires higher-order non-linear effects to capture the correlation $E[\delta \bar{W}_t \delta \bar{Z}_t^2]$. We found that the non-linear component explains about a third of the volatility clustering and is more pronounced on the short-term. The skew of the spot factor combined with the linear spot/vol correlation explains the remaining two thirds.
\item The volatility of volatility is itself volatile, and appears to follow a mean-reverting process with a half-life inferior to a month. The addition of a stochastic time-varying volatility of volatility may be an an interesting approach to generalize the current methodology and integrate additional information provided by the VVIX index (and VIX options).
\item We studied the impact of non-linearities on the dynamics of smiles. As first noted in~\cite{vargas-2013}, the volatility skew generated by non-linear models is in general different from the skewness of the underlying. In the case of the SPX index, for which the linear spot/vol correlation remains the dominant factor, the impact of non-linearities is negligeable and the skew-stickiness ratio defined in~\cite{bergomi-smiledynamicsIV} is practically unchanged. Flatter and/or more-convex volatility smiles, such as the ones on the foreign exchange market, could generate observable differences. 
\item Non-linearities contribute a significant part to the total volatility of the annualised variance, and as such, should be integrated in the pricing, modeling, and hedging of volatility derivatives. The convexity contribution is decreasing more slowly than the discrete sampling impact, but both contributions quickly become smaller than the volatility generated by the remarking of the implied parameters. However, we note that, as long as the correct vega is computed, hedging with spot variance will hedge the different contributions. 
\end{enumerate}

\end{multicols}
 
%
%
%
%
%
%
%
%
%
%
%
%
%
\newpage
\section{Appendix: Proofs}
By considering a small perturbation at first-order in $\theta_\alpha$, the future instantaneous variance $\xi^u_{v}$ observed at a time $v \geq t$ can be approximated from the previous value $\xi^u_t$ by the following formula:
\begin{eqnarray}
\xi^u_{v} = \xi^u_t \times \left( 1 + \sum_{\alpha}{\theta_\alpha \underbrace{\int_{t}^{v}{\omega^{\alpha}(\tau,u,\xi_x) d W^{\alpha}_{\tau}}}_{\text{denoted } \chi_{x,t\to v}^{\alpha,u}}}\right ) = \xi^u_t \times \left( 1 + \sum_{\alpha}{\theta_\alpha \chi_{x,t\to v}^{\alpha,u}}\right ).
\label{eq:perturbeqint}
\end{eqnarray}
where the functions $\omega^\alpha$ are evaluated in the unperturbed state ($\theta_\alpha=0$) with variances frozen at time $x\leq t$ (see~\cite{bergomi-smiledynamicsIV,vargas-2013} for more details). Usually, the frozen time is taken to be either the start-date (i.e. $x=0$), or the current time (i.e. $x=t$). By freezing the variances, the functions $\omega^\alpha$ become deterministic with no stochastic components.
\vspace{2cm}
\subsection{Convexity correction}
\label{app:convexitycorrection}
Under the volatility model defined in Eq.~\ref{eq:var-model}, the value of a VIX future can easily be computed at first-order in the covariance parameters $\Omega_{\alpha,\beta}$. We express the time-$T_1$ term-structure of variance as a perturbation of the term-structure observed at time $t$ by writing $\xi_{T_1}^u=\xi_{t}^u+\psi_t^u(T_1)$ with $d \psi_t^{u}({v}) = \xi_t^u\times \sum_{\alpha}{\theta_\alpha\omega^{\alpha}(v,u,\xi_t) d W^{\alpha}_{v}}
$.\\\
A second-order extension in the perturbation curve $\psi$ leads to:
\begin{eqnarray*}
\mathcal{V}_t^{T_1} &=& E_t\left [\sqrt{\frac{1}{\Delta T}\int_{T_1}^{T_2}{\xi_{T_1}^u du}}\right] = E_t\left[\sqrt{\frac{1}{\Delta T}\int_{T_1}^{T_2}{\left (\xi_{t}^u +\psi_t^u(T_1)\right ) du}}\right ]\\
&\approx& E_t\left [\sqrt{\frac{1}{\Delta T}\int_{T_1}^{T_2}{\xi_{t}^u du}} + \frac{1}{2}\frac{\frac{1}{\Delta T}\int_{T_1}^{T_2}{\psi_t^u(T_1)du}}{\sqrt{\frac{1}{\Delta T}\int_{T_1}^{T_2}{\xi_{t}^u du}}}-\frac{1}{8}\frac{(\frac{1}{\Delta T}\int_{T_1}^{T_2}{\psi_t^u(T_1)du})^2}{(\frac{1}{\Delta T}\int_{T_1}^{T_2}{\xi_{t}^u du})^{\frac{3}{2}}}\right ]\\
&\approx& \mathbb{K}_t^{T_1} - \frac{1}{8}\frac{E_t[(\frac{1}{\Delta T}\int_{T_1}^{T_2}{\psi_t^u(T_1)du})^2]}{(\mathbb{K}_t^{T_1})^{3}}\text{ where } \mathbb{K}_t^{T_1} = \sqrt{\frac{1}{\Delta T}\int_{T_1}^{T_2}{\xi_{t}^u du}}\\
&\approx& \mathbb{K}_t^{T_1} - \frac{1}{8}\frac{E_t[\left ( \sum_{\alpha}{\theta_\alpha \int_t^{T_1}{\delta W^{\alpha}_v \frac{1}{\Delta T}\int_{T_1}^{T_2}{\xi_t^u \omega^{\alpha}(v,u,\xi_t)du}}}\right )^2]}{(\mathbb{K}_t^{T_1})^{3}} \\
&\approx& \mathbb{K}_t^{T_1}\times\left( 1- \underbrace{\frac{1}{8(\mathbb{K}_t^{T_1})^4}\sum_{\alpha,\beta} {\Omega_{\alpha,\beta}  \int_t^{T_1}{dv \frac{1}{\Delta T}\int_{T_1}^{T_2}{\xi_t^u \omega^{\alpha}(v,u,\xi_t)du\times \frac{1}{\Delta T}}\int_{T_1}^{T_2}{\xi_t^u \omega^{\beta}(v,u,\xi_t)du}}}}_{\text{convexity correction}}\right )
\end{eqnarray*}
\newpage
\subsection{Impact of the volatility of volatility on the implied smile}
\label{app:smile}
The presence of vol of vol alters the shape of the implied volatility surface. We introduce a scaling parameter $\lambda$ as $\theta_\alpha \to \lambda \theta_\alpha$ and consider the price of a call option $F_K(\lambda) = E[(S_T-K)^+]$ of strike $K$ in the presence of vol of vol $\lambda\neq 0$. Pricing is achieved under the risk-neutral measure, i.e. $\delta \bar{Z}_t$ and $\delta \bar{W}_t$ follow standard normal distributions. We also neglect the drift component.\\\\
The volatility shift at strike $k$ induced by the presence of vol of vol is:
\begin{eqnarray*}
\delta \sigma (K,t,T) = \sigma^\lambda(K,t,T)-\sigma_{\text{VS}}(t,T) = \frac{F_K(\lambda) - F_K(0)}{\text{Vega}_K}  =  \lambda \frac{F'_K(0)}{\text{Vega}_K}
\end{eqnarray*}
where $\text{Vega}_K$ is the standard Black-Scholes vega. To compute $F_K'(0)$, we follow the same step as in~\cite{vargas-2013}. We express the spot at maturity $S_T$ as a function of an unperturbed state ($\lambda=0$) and a first-order correction:
\begin{eqnarray*}
\log\frac{S_T}{S_t} &=& \sum_u  \log (1+\frac{\delta S_u}{S_u}) \approx \sum_u {\left(\sqrt{\xi_u^u}\delta Z_u-\frac{1}{2}\xi_u^u\delta Z_u^2\right)}\approx \sum_u {\left(\sqrt{\xi_u^u}\delta Z_u-\frac{1}{2}\xi_u^u\delta u\right)}\\
                &\approx& \underbrace{\sum_u {\left(\sqrt{\xi_t^u}\delta Z_u-\frac{1}{2}\xi_t^u\delta u\right)}}_{L_N}
                + \lambda \sum_\alpha \frac{\theta_\alpha}{2}\underbrace{ \sum_u {\left(\sqrt{\xi_t^u}\chi_{t,t\to u}^{\alpha,u} \delta Z_u-\xi_t^u\chi_{t,t\to u}^{\alpha,u}\delta u\right)}}_{\tilde{L}^\alpha_N} \text{ from Eq.~\ref{eq:perturbeqint}}
\end{eqnarray*}
From the above, we have $F(\lambda) = E[(S_t e^{L_N+\lambda \sum_\alpha  \frac{\theta_\alpha}{2}\tilde{L}^\alpha_N}-K)^+]$, so that $F'(0) = \sum_\alpha  \frac{\theta_\alpha}{2} S_t E[\tilde{L}^\alpha_N e^{L_N} \mathbb{1}_{L_N>\log\frac{K}{S_t}}]$. \\\\
For simplicity, we drop the $\alpha$-terms and define $\sigma_u = \sqrt{\xi_t^u \delta t}$ and the moneyness $\mathcal{M}_K = \log{\frac{S_t}{K}} \approx - \frac{dK}{S_t}$. 
The computation of the integral $F_K'(0)$  is painful and computationally intensive. It requires numerous changes of variables and partial integrations. We sketch the proof below. First, we express the integral $F_K'(0)$ as the sum of two integrals:
\begin{eqnarray*}
F_K'(0) &=& S_t E\left[{ \tilde{L}_N e^{L_N}\mathbb{1}_{L_N>\log\frac{K}{S_t}}}\right]\\
                                &=& S_t E\left[{ \tilde{L}_{N-1} e^{L_{N-1}} \Phi(\frac{\mathcal{M}_K + L_{N-1}+\frac{1}{2}\sum_{i=N}^N{\sigma_i^2}}{\sqrt{\sum_{i=N}^N{\sigma_i^2}}})}\right] 
                                 + \sigma_N S_t E\left[{ e^{L_{N-1}}\chi_{N-1}^{N}  \frac{1}{\sqrt{2\pi}} e^{-\frac{1}{2}(\frac{\mathcal{M}_K + L_{N-1}+\frac{1}{2}\sum_{i=N}^N{\sigma_i^2}}{\sqrt{\sum_{i=N}^N{\sigma_i^2}}})^2} }\right] \\
                                &=& S_t \times [I(N-1) + \sigma_N J(N-1)]
\end{eqnarray*}
We then make use of the following equality $\int{ \frac{1}{\sqrt{2\pi}} e^{-\frac{1}{2}x^2} \Phi(ax + b)} = \Phi(\frac{b}{\sqrt{1+a^2}})$, to derive the recursive equality:
\[
I(k-1) = I(k-2)+  \frac{\sigma_{k-1}^2}{\sqrt{\sum_{i=k-1}^N{\sigma_i^2}}}  J(k-2)
\]
The integral $J(k-1)$ can be computed as:
\begin{eqnarray*}
J(k-1) &=&  e^{-\mathcal{M}_K} \sum_{u<k}{\lambda_u^k \sqrt{\frac{\sum_{i=k}^N{\sigma_i^2}}{\sum_{i\neq u }{\sigma_i^2}}} E\left[{ \delta \bar{W}_u \frac{1}{\sqrt{2\pi}}  e^{-\frac{1}{2}\frac{(\sigma_u \epsilon_u + \mathcal{M}_K - \frac{1}{2}\sum{\sigma_i^2})^2}{\sum_{i\neq u }{\sigma_i^2}}  }}\right]}\\
&=& \frac{1}{\sqrt{2\pi}}e^{-\frac{1}{2}\frac{(\mathcal{M}_K + \frac{1}{2}\sum{\sigma_i^2})^2}{\sum{\sigma_i^2}}}\sum_{u<k}{\lambda_u^k \sqrt{\frac{\sum_{i=k}^N{\sigma_i^2}}{\sum{\sigma_i^2}}}  E\left[{f_\alpha(\frac{\sigma_u}{\sum{\sigma_i^2}}  (\frac{1}{2}\sum{\sigma_i^2}-\mathcal{M}_K)+ \sqrt{\frac{\sum_{i\neq u }{\sigma_i^2}}{\sum{\sigma_i^2}}} U)  }\right]}
\end{eqnarray*}
We denote $\mathbb{VaR}$ the total Black-Scholes variance $\mathbb{VaR} = \sum_{u=0}^N \sigma^2_u \approx \int_t^T{\xi_t^u du}$. Putting everything together, we find that:
\begin{eqnarray*}
\frac{F_K'(0)}{\text{Vega}_K} &=& \sum_\alpha {\frac{\theta_\alpha }{2\sqrt{(T-t)\mathbb{VaR}}}\sum_{u}{\left [\xi_t^u \delta u \sum_{v<u}{\left( \delta v \omega^\alpha(u,v)  E\left[\frac{1}{\sqrt{\delta t}}{f_\alpha(\frac{\sqrt{\xi_t^u\delta u}}{\mathbb{VaR}}  (\frac{1}{2}\mathbb{VaR}-\mathcal{M}_K)+ \sqrt{\frac{\mathbb{VaR}-\xi_t^u\delta u}{\mathbb{VaR}}} U)  }\right]  \right ) } \right ]}}
\end{eqnarray*}
\newpage
\subsection{Volatility of the annualised variance}
\label{app:variance}
In order to derive the total variance of the annualized variance $E_{T_1}[\frac{1}{\Delta T}\int_{T_1}^{T_2}{(\frac{\delta \mathbb{V}_u}{\mathbb{V}_u})^2}]$  from eq.~\ref{eq:vardecompo}, we make the assumption of a relatively flat-term structure of variance. We assume that the differences between $\mathbb{V}_t$, implied variance $\mathbb{I}_t$ and instantaneous variance $\xi_t^u$ for $u\geq t$ should be of second-order (compared to the different integral terms involving the covariance parameters, kurtosis, and correlation terms). Although this approximation is rarely verified exactly, it does bring the advantage of obtaining an accurate closed-form solution.\\\\
Under our assumption, the following approximation can be derived
\[
\int_t^{T_2}{d\xi_t^u du}= \sum_{\alpha=1}^{n} {\theta_\alpha \int_t^{T_2}{\xi_t^u e^{-k_\alpha (T_2-u)}du } d W^{\alpha}_t}\approx \mathbb{I}_t \sum_{\alpha=1}^{n} {\theta_\alpha  g_\alpha(T_2-t) d W^{\alpha}_t} \text{ with } g_\alpha(T_2-t) = \frac{1-e^{-k_\alpha (T_2-t)}}{k_\alpha(T_2-t)},
\]
which applied to Eq.~\ref{eq:vardecompo} leads to the variation of the mark-to-market
\begin{eqnarray*}
\delta \mathbb{V}_t\approx \frac{1}{\Delta T}\xi^t_t\left(\delta Z_t^2-  \delta t \right) + \frac{T_2-t}{\Delta T}\mathbb{I}_t \sum {\theta_\alpha g_\alpha(T_2-t) d W^{\alpha}_t}
\end{eqnarray*}
From there it is easy to compute the total variance. Our flat-term structure assumption means that the different volatility the ratios (see below) can be neglected without much impact on the final solution.
\begin{eqnarray*}
E_{T_1}[\frac{1}{\Delta T}\int_{T_1}^{T_2}{(\frac{\delta \mathbb{V}_u}{\mathbb{V}_u})^2}] &\approx& \frac{1}{\Delta T^3}\int_{T_1}^{T_2}{E_{T_1}[\cancel{(\frac{\xi^u_u}{\mathbb{V}_u})^2}(\delta Z_u^2-  \delta u)^2]}+ \frac{1}{\Delta T^3} \sum {\frac{\Omega_{\alpha,\beta}}{k_\alpha k_\beta} \int_{T_1}^{T_2}{E_{T_1}[\cancel{\frac{\mathbb{I}_u}{\mathbb{V}_u^2}}] (1-e^{-k_\alpha (T_2-u)})(1-e^{-k_\beta (T_2-u)})du}}\\
&& + \frac{2}{\Delta T^3} \sum {\frac{\theta_\alpha}{k_\alpha} \int_{T_1}^{T_2}{E_{T_1}[\cancel{(\frac{\xi_u^u \mathbb{I}_u}{\mathbb{V}_u})^2} (\delta Z_u^2-  \delta u)d W^{\alpha}_u ] (1-e^{-k_\alpha (T_2-u)})}}\\
&\approx& \frac{\kappa + 2}{N\Delta T} + \sum_{\alpha,\beta}{\Omega_{\alpha,\beta} \underbrace{\frac{1}{k_\alpha k_\beta \Delta T}(\frac{1}{\Delta T} - \frac{1-e^{-k_\alpha \Delta T}}{k_\alpha\Delta T^2}-\frac{1-e^{-k_\beta \Delta T}}{k_\beta\Delta T^2}+\frac{1-e^{-(k_\alpha+k_\beta) \Delta T}}{(k_\alpha+k_\beta)\Delta T^2})}_{l(k_\alpha,k_\beta,\Delta T)}} \\
&&+ 2\sqrt{\frac{\kappa + 2}{N\Delta T}}\sum_{\alpha}{\rho_{\text{shocks}}^{\alpha} \theta_\alpha \underbrace{(\frac{1}{k_\alpha \Delta T}-\frac{1-e^{-k_\alpha \Delta T}}{(k_\alpha\Delta T)^2})}_{h(k_\alpha \Delta T)}}
\end{eqnarray*}
\newpage
\bibliography{80cap}
\bibliographystyle{abbrv}

\end{document}